\documentclass[12pt]{article} %
\RequirePackage[colorlinks,citecolor=blue,urlcolor=blue,filecolor=red]{hyperref}
\usepackage[utf8]{inputenc} %
\usepackage{geometry} %
\usepackage{graphicx} %

\usepackage{booktabs} %
\usepackage{tabularx} %
\usepackage{array} %
\usepackage{paralist} %
\usepackage{verbatim} %
\usepackage{subcaption} %
\usepackage{floatrow}
\usepackage{amsbsy}
\usepackage{amsmath}
\usepackage{amssymb}
\usepackage{float}
\usepackage[toc,page,title]{appendix}
\usepackage[]{natbib}
\usepackage{lineno}
\usepackage{mathrsfs}
\usepackage{todonotes}
\usepackage{setspace}
\usepackage{natbib}
\usepackage{multirow}
\usepackage{textcomp}
\usepackage{makecell}
\usepackage{blkarray}

\usepackage{multibib}
\newcites{si}{Supplemental References}%

\usepackage{pfa_spacing}

\input{operators.tex}

\def\siSecFirst{Supplemental Information (SI) Section}
\def\siSec{SI Section}
\def\siTab{SI Table}

\def\spacingset#1{\renewcommand{\baselinestretch}%
	{#1}\small\normalsize} \spacingset{1}

\newif\ifsiappended
\newcommand*\siref[1]{\ifsiappended\ref{#1}\else\ref*{#1}\fi}
\siappendedfalse

\newif\ifblinded
\newcommand{\blind}[1]{\ifblinded{[blinded]}\else#1\fi}

\blindedfalse %

\begin{document}

\pagenumbering{gobble}
\begin{flushright}
	Version dated: \today
\end{flushright}

\medskip
\begin{center}
	\centerline{\Large \bf Principled, practical, flexible, fast:}
	\centerline{\Large \bf a new approach to phylogenetic factor analysis}
	\bigskip

\blind{\noindent{\normalsize \sc
	Gabriel W.~Hassler$^1$, \\
	Brigida Gallone$^2$, \\
	Leandro Aristide$^3$, \\
	William L.~Allen$^4$, \\
	Max R.~Tolkoff$^5$, \\
	Andrew J.~Holbrook$^5$, \\
	Guy Baele$^6$, \\
	Philippe Lemey$^6$ \\
	and Marc A.~Suchard$^{1, 5, 7}$
}

\vspace{1em}

\noindent {\small
	\it $^1$Department of Computational Medicine, David Geffen School of Medicine at UCLA, University of California,
	Los Angeles, United States \\
	\it $^2$VIB–KU Leuven Center for Microbiology, Leuven, Belgium\\
	\it $^3$Ecole Normale Superieure Paris Sciences et Lettres Research University, Institut de Biologie de l’Ecole Normale Superieure, Paris, France\\
	\it $^4$Department of Biosciences, Swansea University, Swansea, United Kingdom \\
	\it $^5$Department of Biostatistics, Jonathan and Karin Fielding School of Public Health, University
	of California, Los Angeles, United States} \\
	\it $^6$Department of Microbiology and Immunology, Rega Institute, KU Leuven, Leuven, Belgium \\
	\it $^7$Department of Human Genetics, David Geffen School of Medicine at UCLA, Universtiy of California,
Los Angeles, United States}

\vfill

\end{center}

\input{pfa_vars.sty} %
\input{pfa_results.sty} %

\paragraph{Abstract}
Biological phenotypes are products of complex evolutionary processes in which selective forces influence multiple biological trait measurements in unknown ways.
Phylogenetic factor analysis disentangles these relationships across the evolutionary history of a group of organisms.
Scientists seeking to employ this modeling framework confront numerous modeling and implementation decisions, the details of which pose computational and replicability challenges.
General and impactful community employment requires a data scientific analysis plan that balances flexibility, speed and ease of use, while minimizing model and algorithm tuning.
Even in the presence of non-trivial phylogenetic model constraints, we show that one may analytically address latent factor uncertainty in a way that (a) aids model flexibility, (b) accelerates computation (by as much as 500-fold) and (c) decreases required tuning.
We further present practical guidance on inference and modeling decisions as well as diagnosing and solving common problems in these analyses.
We codify this analysis plan in an automated pipeline that distills the potentially overwhelming array of modeling decisions into a small handful of (typically binary) choices.
We demonstrate the utility of these methods and analysis plan in four real-world problems of varying scales.

\vspace{1em}


\clearpage
\pagenumbering{arabic}

\section{Introduction} \label{sec:intro}
Biological phenotypes are the result of numerous evolutionary forces acting in complex and often conflicting ways throughout an organism's evolutionary history.
Phylogenetic comparative methods seek to untangle this web of selective pressures and elucidate the forces that have shaped organisms over time.
As implied by their name, these methods compare phenotypes across numerous biological taxa connected by a phylogenetic tree that captures their shared evolutionary history.
Accounting for shared evolutionary history via the phylogeny is necessary to avoid biased inference, as this shared history implies phenotypes are non-independent across taxa.
Statistical models that inappropriately ignore this dependence can identify spurious associations between phenotypes \citep[][]{felsenstein1985phylogenies}.
However, accounting for these relationships between taxa poses challenges to statistical inference.

Starting with \citet{felsenstein1985phylogenies}, there has been much work developing computationally efficient phylogenetic comparative methods \citep[see][]{rohlf2001comparative, revell2008testing, Pybus2012, ho2014linear}.
While methods development has typically focused on scaling inference to large trees, these methods struggle to accommodate data with a large number of traits or high-dimensional phenotypes.
Most approaches scale quadratically or cubically with the number of traits, making inference intractable as the number of traits increases.
Additionally, methods that estimate the evolutionary correlation structure between traits are difficult to interpret for data sets with high-dimensional phenotypes, as the number of pairwise correlations requiring interpretation scales quadratically with the number of traits.

Linear dimension reduction is a common strategy to ease the computational and interpretative burden of high-dimensional data.
Phylogenetic principle components analysis \citep[pPCA, ][]{revell2009size} is one solution that constructs a low-dimensional, phylogenetically-informed summary of the relationships between traits.
This method, however, is not likelihood-based, making uncertainty difficult to assess and integration into larger, likelihood-based models challenging.
More recently, several distance-based methods have been developed by \citet{adams2014generalized, adams2014method, adams2014quantifying}.
While these methods efficiently scale to high-dimensional data sets, each is specific to a particular problem (i.e.~phylogenetic signal, phylogenetic regression and evolutionary rates, respectively) and not easily generalizable.
Additionally, neither the work of \citet{revell2009size} nor \citet{adams2014generalized, adams2014method, adams2014quantifying} readily accommodate missing data, a common scourge in many relevant data sets.
Phylogenetic factor analysis \citep[PFA,][]{tolkoff2017phylogenetic} adapts the latent factor model of \citet{aguilar2000bayesian} to the phylogenetic context.
Like pPCA, PFA is a linear dimension reduction approach that assumes the $\ntraits$-dimensional data arise from $\nfac$ latent factors that evolve independently along a phylogenetic tree.
While PFA accommodates missing data and provides a more generalizable likelihood-based alternative to pPCA, the Bayesian inference regime proposed by \citet{tolkoff2017phylogenetic} scales quadratically with the number of taxa and is intractable for large trees.

Finally, \citet{clavel2019penalized} propose a penalized-regression framework for studying high-dimensional phenotypes.
However, this procedure is designed for data sets with few taxa and does not scale well to large trees.

\newcommand{\traitFormat}[1]{#1}
\def\traitOne{\traitFormat{A}}
\def\traitTwo{\traitFormat{B}}
\def\traitThree{\traitFormat{C}}
\newcommand{\traits}[3]{#1, #2, #3}
\def\traitsOne{\traits{\traitOne}{\traitTwo}{\traitThree}}
\def\traitsTwo{\traits{\traitTwo}{\traitOne}{\traitThree}}

\newcommand{\facOneOrder}[3]{captures rank-one relationships of trait #1 with traits #2 and #3}
\newcommand{\facTwoOrder}[3]{captures relationships between traits #2 and #3 independent of #1}
\def\orderRules{\cmidrule(l{2pt}r{2pt}){2-2}\cmidrule(l{2pt}r{2pt}){3-3}} %
\begin{table}
	\centering
	\caption{Example of how the ordering of three hypothetical traits (\traitOne, \traitTwo~and \traitThree) influences results in a simple two-factor model under the assumptions made by \citet{tolkoff2017phylogenetic}.}
	\label{tb:facOrder}
	\begin{tabularx}{\linewidth}{r X X}\toprule%
		& trait order 1: \traitsOne & trait order 2: \traitsTwo\\ \orderRules
		first factor & \facOneOrder{A}{B}{C} & \facOneOrder{B}{A}{C}\\ \orderRules
		second factor & \facTwoOrder{A}{B}{C} & \facTwoOrder{B}{A}{C} \\
		\bottomrule
	\end{tabularx}
\end{table}

We propose two new PFA inference regimes that each scale linearly with both the number of traits $\ntraits$ and the number of taxa $\ntaxa$.
While \citet{tolkoff2017phylogenetic} rely on data augmentation, our new methods rely on a novel likelihood-calculation algorithm that analytically integrates out the latent factors.
We also address two other shortcomings of PFA and latent factor models generally.
First, \citet{tolkoff2017phylogenetic} constrain the factor loadings matrix to be upper triangular, which induces an implicit ordering to the phenotypes. %
Specifically, the first trait is influenced only by the first factor, the second trait is influenced only by the first two factors, etc.~until the $\nfac^{\text{th}}$ trait and beyond which are influenced by all $\nfac$ factors (see Table \ref{tb:facOrder} for an example).
As justifying a specific ordering of the phenotypes \emph{a priori} can be difficult, we extend an alternative constraint proposed by \citet{holbrook2016bayesian} that eliminates such ordering.
Second, a common challenge in exploratory factor analysis generally is determining an appropriate number of factors.
As such, we implement a cross-validation model selection procedure that identifies the number of factors that confers the best predictive performance.

To facilitate use among researchers seeking to employ these methods, we develop an analysis plan with practical guidance on the most significant modeling and inference decisions.
We codify this plan in the Julia package PhylogeneticFactorAnalysis.jl, which uses relatively simple instructions to automatically perform model selection and run more complex analyses in the Bayesian phylogenetic inference software BEAST \citep{BEAST}.

\section{Phylogenetic Latent Factor Model}
We approach inference from a Bayesian perspective and propose two statistical models which share a likelihood but have distinct priors.
As we discuss below, each model has advantages under different circumstances, and allowing researchers to choose a model (with our guidance) offers maximum flexibility while keeping modeling decisions to a minimum.

\subsection{Likelihood}

Both statistical models share the same latent factor likelihood introduced by \citet{tolkoff2017phylogenetic}.
This likelihood assumes the $\ntaxa\times\ntraits$ trait data $\data = \left(\datarow{1}\too\datarow{\ntaxa}\right)\trans$ arise from $\ntaxa\times\nfac$ latent factors $\fac = \left(\facrow{1}\too\facrow{\ntaxa}\right)\trans$ via the linear transformation $\data = \fac\load + \error$, where $\load$ is a $\nfac\times\ntraits$ loadings matrix that must be inferred and $\error \sim \mndist{\bzero}{\I{\ntaxa}}{\errprec\inv}$ is matrix-normally distributed with mean $\bzero$, between row variance $\I{\ntaxa}$ and diagonal between column precision $\errprec = \diagOp{\errprecel{1}\too\errprecel{\ntraits}}$.
The latent factors $\fac$ arise from $\nfac$ independent Brownian diffusion processes on the phylogenetic tree $\tree$.
This tree $\tree$ is a directed acyclic graph with degree-two root node $\node{\nroot}$, degree-three internal nodes $\left\{\node{\ntaxa + 1}\too\node{2\ntaxa - 2}\right\}$ and degree-one terminal nodes $\left\{\node{1}\too\node{\ntaxa}\right\}$.
Under the Brownian diffusion model, %
all internal and tip factors are distributed as $\facrow{\taxaITwo} \sim \ndist{\facrow{\parent{\taxaITwo}}}{\blen{\taxaITwo}\I{\nfac}}$, where $\facrow{\parent{\taxaITwo}}$ are the factors of the parent of node $\node{\taxaITwo}$ and $\blen{\taxaITwo}$ is the distance (time) between nodes $\node{\parent{\taxaITwo}}$ and $\node{\taxaITwo}$.
Following from \citet{Pybus2012}, we assume the ancestral root traits
$\facrow{\nroot}\sim\ndist{\rootmean}{\frac{1}{\pss}\I{\nfac}}$,
where $\pss$ is some (typically small) predetermined prior sample size.
This construction implies the tip factors are jointly distributed as
$\fac\sim\mndist{\onevec{\ntaxa}\rootmean\trans}{\treevar + \frac{1}{\pss}\onemat{\ntaxa}}{\I{\nfac}}$,
where $\onevec{\ntaxa}$ is a $\ntaxa$-vector of ones, $\onemat{\ntaxa} = \onevec{\ntaxa}\onevec{\ntaxa}\trans$ and $\treevar$ is the standard variance-covariance (VCV) representation of the phylogeny $\tree$. %
Specifically, the diagonal elements $\treevarel{\taxaI\taxaI}$ are the sum of the edge lengths connecting $\node{\taxaI}$ to the root $\node{\nroot}$.
The off-diagonal elements $\treevarel{\taxaI\taxaITwo}$ are the total amount of shared evolutionary history or time from the most recent common ancestor of $\node{\taxaI}$ and $\node{\taxaITwo}$ to the root node $\node{\nroot}$.

Given this model, the vectorized data $\vecOp{\data}$ are multivariate normally distributed as
\begin{equation}\label{eq:naiveLikelihood}
    \cdist{\vecOp{\data}}{\load, \errprec, \tree} \sim \ndist{\vecOp{\onevec{\ntaxa}\rootmean\trans}}{\load\trans\load\kron\left[\treevar + \frac{1}{\pss}\onemat{\ntaxa}\right] + \errprec\inv\kron\I{\ntaxa}},
\end{equation}
where $\kron$ is the Kronecker product operator.
Computing the likelihood in this form, however, requires inverting the $\ntaxa\ntraits\times\ntaxa\ntraits$ dimensional variance matrix, which has computational complexity $\bigo{\ntaxa^3\ntraits^3}$.
\citet{tolkoff2017phylogenetic} avoid this by treating the latent factors $\fac$ as model parameters that they integrate out via Markov chain Monte Carlo (MCMC) simulation.
This augmented likelihood $\cdens{\data, \fac}{\load, \errprec, \tree} = \cdens{\data}{\load, \errprec, \fac}\cdens{\fac}{\tree}$ is far easier to compute, but sampling from the full conditional distribution of $\fac$ as proposed by \citet{tolkoff2017phylogenetic} scales quadratically with the size of the phylogenetic tree and is intractable for big-$\ntaxa$.

\subsubsection{Fast Likelihood Calculation} \label{sec:fastLikelihood}

To avoid costly data augmentation, we adapt the likelihood-computation algorithm independently developed by \citet{bastide2018inference}, \citet{mitov2019fast} and \citet{hassler2020inferring}.
This algorithm analytically integrates out latent traits (in our case factors) and computes the likelihood $\cdens{\data}{\load, \errprec, \tree}$ in $\bigo{\ntaxa\nfac^3}$ via a post-order traversal of the tree.
This procedure naturally accommodates missing data assuming an ignorable missing data mechanism \citep{rubin1976inference}.
Let $\obsdata = \left(\obsdatarow{1}, \hdots, \obsdatarow{\ntaxa}\right)\trans$ be the $\ntaxa\times\ntraits$ matrix of observed data, where all missing measurements in $\data$ have been replaced with $0$'s.
This post-order algorithm requires that one can compute the partial mean $\pmean{\taxaI}$, precision $\pprec{\taxaI}$ and remainder $\prem{\taxaI}$ such that
\begin{equation}
    \begin{aligned}
        \cdens{\obsdatarow{\taxaI}}{\facrow{\taxaI}, \load, \errprec} &= \prem{\taxaI}\sndens{\facrow{\taxaI}}{\pmean{\taxaI}}{\pprec{\taxaI}},\text{ where} \\
        \sndens{\arbarg}{\arbmean}{\arbprec} &= \left(2\pi\right)^{-\rankscr{\arbprec}/2}\sdet{\arbprec}^{1/2}\expOp{-\frac{1}{2}\left(\arbarg - \arbmean\right)\trans\arbprec\left(\arbarg - \arbmean\right)},
    \end{aligned}
\end{equation}
$\rank{\arbprec}$ is the number of non-zero singular values of $\arbprec$ and $\sdet{\arbprec}$ is the product of the non-zero singular values of $\arbprec$.
We also define the indicator matrices $\traitmismat{\taxaI} = \diagOp{\misind{\taxaI 1}\too\misind{\taxaI\ntraits}}$ where $\misind{\taxaI\traitI} = 1$ if $\datum{\taxaI\traitI}$ is observed and $\misind{\taxaI\traitI} = 0$ if it is missing.
Finally, we define $\ntraitobs{\taxaI} = \sum_{\traitI = 1}^{\ntraits} \misind{\taxaI\traitI}$ as the number of observed traits for taxon $\taxaI$.

In the context of PFA, we calculate
\begin{equation} \label{eq:factorPartials}
    \begin{aligned}
        \log\cdens{\obsdatarow{\taxaI}}{\facrow{\taxaI}, \load, \errprec} &= -\frac{\rank{\misprec{\taxaI}}}{2}\log2\pi + \frac{1}{2}\log\sdet{\misprec{\taxaI}}\\
        &\hspace{0.5in} - \frac{1}{2}\left(\obsdatarow{\taxaI} - \load\trans\facrow{\taxaI}\right)\trans\misprec{\taxaI} \left(\obsdatarow{\taxaI} - \load\trans\facrow{\taxaI}\right) \\
        &= \log\prem{\taxaI} + \log\sndens{\facrow{\taxaI}}{\pmean{\taxaI}}{\pprec{\taxaI}},\text{ where }
    \end{aligned}
\end{equation}
the precision $\pprec{\taxaI} = \load\misprec{\taxaI}\load\trans$, the mean $\pmean{\taxaI}$ is a solution to $\pprec{\taxaI}\pmean{\taxaI} = \load\trans\misprec{\taxaI}\obsdatarow{\taxaI}$ and
\begin{equation}
    \begin{aligned}
        \log\prem{\taxaI} = -\frac{\ntraitobs{\taxaI} - \rank{\pprec{\taxaI}}}{2}&\log 2\pi + \frac{1}{2}\left(\sum_{\traitI = 1}^{\ntraits} \misind{\taxaI\traitI}\log \errprecel{\traitI} - \log\sdet{\pprec{\taxaI}}\right)\\
         &- \frac{1}{2}\left[\obsdatarow{\taxaI}\trans\misprec{\taxaI}\obsdatarow{\taxaI}  - \pmean{\taxaI}\trans\pprec{\taxaI}\pmean{\taxaI}\right].
    \end{aligned}
\end{equation}
See \siSecFirst~\siref{ap:partials} for detailed calculations.
As $\errprec$ is diagonal, computing all $\pprec{\taxaI}$ has complexity $\bigo{\ntaxa\ntraits\nfac^2}$, which dominates the computation time for these operations.

After computing $\pmean{\taxaI}$, $\pprec{\taxaI}$ and $\prem{\taxaI}$, the \citet{hassler2020inferring} algorithm requires minor modification to compute the likelihood $\cdens{\obsdata}{\load, \errprec, \tree}$ in $\bigo{\ntaxa\nfac^3}$ additional time.
Specifically, $\pprec{\taxaI}$ may not be invertible via the special inverse defined in \citet{hassler2020inferring}.
\siSec~\siref{ap:inverse} offers an alternative approach that avoids this inversion via the continuously rediscovered identity $
\left(\matA + \matB\right)\inv = \matA\inv - \matA\inv\left(\I{} + \matB\matA\inv\right)\inv\matB\matA\inv
$ for conformable square matrices $\matA$ and $\matB$ \citep[][]{henderson1959estimation, henderson1981on}.
We also utilize a more numerically stable modification of this post-order algorithm proposed by \citet{bastide2020efficient}.

\subsubsection{Loadings Identifiability}\label{sec:identifiability}
\def\loadalt{\left(\orthog\load\right)}
A major challenge in latent factor models generally is the identifiability of the loadings matrix $\load$ \citep[see][]{shapiro1985identifiability}.
This lack of identifiability stems from the fact that the likelihood as defined in Equation \ref{eq:naiveLikelihood}
depends only on $\load\trans\load$ rather than $\load$ itself.
As such, for any $\nfac \times \nfac$ orthonormal matrix $\orthog$ (i.e.~$\orthog\trans\orthog = \I{\nfac}$), $\cdens{\data}{\load,\hdots} = \cdens{\data}{\orthog\load,\hdots}$ because ${\loadalt}\trans\loadalt = \load\trans\load$. %
This identifiability problem inspires our choice of priors below.

\subsection{Priors} \label{sec:priors}
We assume the diagonal precisions $\errprecel{\traitI} \sim \gammadist{\errshape}{\errrate}$ for $\traitItoo$ (shape/rate parameterization).
For the loadings $\load = \{\loadij\}$, we propose two different priors.
Each prior on $\load$ admits a different inference regime for sampling from $\load$ which in turn have their own strengths and weaknesses that we discuss in Section \ref{sec:inference}.

\subsubsection{Independent Gaussian Priors on the Loadings \texorpdfstring{$\load$}{L}} \label{sec:iid}

The standard assumption in Bayesian latent factor models is that each element of the loadings $\loadel{\facI\traitI}\stackrel{\text{\tiny i.i.d.}}{\sim}\ndist{0}{\loadVariid}$, where typically $\loadVariid = 1$. %
As this prior is also invariant with respect to orthogonal rotations, additional constraints are required for posterior identifiability.
One solution is to enforce structured sparsity in the model, which typically involves fixing all elements of $\load$ below the diagonal to 0 \citep{geweke1996measuring, aguilar2000bayesian}.
This approach solves the identifiability problem, but it induces an implicit ordering to the data (see Table \ref{tb:facOrder}).
While this ordering may be well-informed in some cases, there is typically no principled way to choose such an ordering \emph{a priori}.

An alternative to the sparsity constraint is to assume that the loadings matrix has rows that 1) are orthogonal and 2) have decreasing norms \citep{holbrook2016bayesian}.
This constraint does not require any \emph{a priori} ordering of the traits.
However, it does require sampling from the space of orthogonal matrices, which is a notoriously challenging problem \citep[see][]{hoff2009simulation, byrne2013geodesic,jauch2020monte, pourzanjani2021bayesian}. %
We address this challenge via post-processing in Section \ref{sec:svdProcessing}. %

\subsubsection{Orthogonal Shrinkage Prior} \label{sec:shrink}
While post-processing to orthogonality is often sufficient, we find in practice that the loadings may be only loosely identifiable with this procedure in small-$\ntaxa$ problems.
As such, we seek an alternative prior that enforces the orthogonality constraint directly.
Following from \citet{holbrook2017bayesian}, we decompose the loadings $\load = \loadScale\loadOrtho$ where $\loadScale$ is a $\nfac\times\nfac$ diagonal matrix whose diagonals have descending absolute values and $\loadOrtho$ is a $\nfac\times\ntraits$ orthonormal matrix (i.e.~$\loadOrtho\loadOrtho\trans = \I{\nfac}$).
We assume $\loadOrtho\trans$ is uniformly distributed over the Stiefel manifold $\stiefel{\nfac}{\ntraits}$ (i.e.~the space of $\ntraits\times\nfac$ orthonormal matrices). %
For the scale component $\loadScale = \diagOp{\loadScaleEl{1}\too\loadScaleEl{\nfac}}$ we assume a multiplicative gamma prior inspired by \citet{bhattacharya2011sparse}:
\def\gshrinki{\gshrinkel{\facI}}
\def\multi{\shrinkMult{\facITwo}}
\def\loadScalei{\loadScaleEl{\facI}}
\begin{equation}
\begin{aligned}
\loadScalei &\sim\nDist{0}{\gshrinki^{-1}} \text{ for } \facI = 1\too\nfac, \text{ where }\\
\gshrinki &= \prod_{1}^\facI \multi \text{ and}\\
\multi &\sim \gammaDist{\multshape{\facITwo}}{\multrate{\facITwo}} \text{ for } \facITwo = 1\too\nfac.
\end{aligned}
\end{equation}
For $\facITwo > 1$, we constrain the prior shape $\multshape{\facITwo}$ and rate $\multrate{\facITwo}$ such that $\multshape{\facITwo} > \multrate{\facITwo}$ (i.e.~$\expectOp{\shrinkMult{\facITwo}} > 1$).
This constraint implies that the $\gshrinkel{\facI}$ are (stochastically) increasing with $\facI$.

This prior induces posterior identifiability, as it is not invariant under rotations of the loadings. %
It also provides a data-driven approach for determining the number of factors.
The stochastically increasing $\gshrinki$ place increasing prior mass near 0 for each successive row of the loadings.
For large enough $\nfac$, there will be some $\nfaceff < \nfac$ such that for $\facI > \nfaceff$ each element $\loadij$ has high posterior density near zero.
In this case only the first $\nfaceff$ factors are significantly involved in the process of interest, and the remaining can be ignored.

\def\scaleThreshold{\alpha}
In some cases, particularly when $\nfac$ is relatively large (i.e.~$>5$), we find that this prior does not induce sufficient identifiability in practice.
For these cases, we multiply the joint prior on $\loadScale$ by an indicator function $\Indicator{\absOp{\loadScaleEl{\facI}} < \scaleThreshold\absOp{\loadScaleEl{\facI - 1}} \text{ for }\facI = 2\too\nfac}$.
Setting $\alpha < 1$ forces spacing between the diagonals of $\loadScale$, which results in more identifiable posteriors.

\section{Inference} \label{sec:inference}
Our Bayesian inference regime seeks to approximate the posterior distribution of the parameters of scientific interest via MCMC simulation.
We typically use molecular sequence data $\seqdata$ to simultaneously infer the factor model parameters and phylogenetic tree by approximating
\begin{equation}
\cdens{\load,\errprec,\tree}{\obsdata,\seqdata} \propto \cdens{\obsdata}{\load, \errprec, \tree}\dens{\tree,\seqdata}\dens{\load}\dens{\errprec},
\end{equation}
where $\dens{\tree,\seqdata}$ is developed elsewhere \citep[see][]{BEAST}. %
For cases where we lack sequence data or $\tree$ is too large to infer efficiently, we simply place a degenerate prior on $\tree$.
Generally, we employ a random-scan Metropolis-within-Gibbs \citep{liu1995covariance} approach to inference where at each step in the Markov chain we randomly select a (set of) parameter(s) to update conditioning on the current state of all other parameters in the chain.

\subsection{Loadings Under the i.i.d.~Gaussian Prior}

We propose two different samplers to draw from the full conditional distribution of the loadings $\load$ under the i.i.d.~Gaussian prior from Section \ref{sec:iid}.
The first relies on the Gibbs sampler used by \citet{tolkoff2017phylogenetic}, where we sample from $\cdist{\load}{\obsdata,\fac, \errprec}$.
The second avoids data augmentation and can sample directly from the full conditional distribution $\cdist{\load}{\obsdata, \errprec, \tree}$ without conditioning on the latent factors $\fac$.

\subsubsection{Gibbs Sampler with Data Augmentation}
\label{sec:gibbsLoadings}

\citet{tolkoff2017phylogenetic} use the conjugate Gibbs sampler of \citet{lopes2004bayesian} to sample from $\cdist{\load}{\obsdata, \fac, \errprec}$.
As this sampler conditions on the latent factors $\fac$, \citet{tolkoff2017phylogenetic} simultaneously infer the factors by sequentially drawing from $\cdist{\facrow{\taxaI}}{\facnot{\taxaI}, \obsdata, \load, \errprec, \tree}$ for $\taxaItoo$, where $\facnot{\taxaI}$ represents all factors except $\facrow{\taxaI}$.
As sampling $\facrow{\taxaI}$ for all $\ntaxa$ taxa requires $\bigo{\ntaxa^2\nfac^2}$ work, this procedure quickly becomes intractable with increasing taxa.

Rather than relying on this per-taxon sampling scheme, we employ the pre-order data augmentation algorithm of \citet{hassler2020inferring} that %
uses statistics from the post-order likelihood computation to draw jointly from $\cdist{\fac}{\obsdata, \load,\errprec, \tree}$ in $\bigo{\ntaxa\nfac^3}$ via a single pre-order traversal of the tree (see \siSec~\siref{ap:preOrder} for details).
After sampling from $\cdist{\fac}{\obsdata, \load, \errprec, \tree}$, we can draw directly from $\cdist{\load}{\obsdata, \fac, \errprec}$ using the procedure developed by \citet{lopes2004bayesian} with computational complexity $\bigo{\ntaxa\nfac^2\ntraits}$ (see \siSec~\siref{ap:loadGibbs} for details).

\subsubsection{Hamiltonian Monte Carlo Sampler} \label{sec:hmcLoadings}

We also propose an alternative Hamiltonian Monte Carlo \citep[HMC;][]{HMC} sampler for the loadings that does not require data augmentation.
Intuitively, HMC (a form of MCMC) treats parameter values as the position of a particle in a landscape informed by the posterior distribution.
Parameter proposals are the end-point of a trajectory initiated by ``kicking" the particle and allowing it to traverse this landscape according to Hamiltonian dynamics for a pre-determined amount of time.
As the parameter trajectories are informed by the geometry of the posterior, HMC tends to propose parameter updates that are both relatively far away from the current position and have high acceptance probabilities.

While we cannot compute these continuous trajectories analytically, we approximate the solution to these ordinary differential equations using first-order numerical methods.
Each trajectory approximation, however, requires numerous gradient calculations, and we must efficiently compute the gradient
$\grad{\logcdens{\load}{\obsdata, \errprec, \tree}}{\load} = \grad{\logcdens{\obsdata}{\load, \errprec,\tree}}{\load} +  \grad{\logdens{\load}}{\load}$
to effectively employ HMC to update the loadings $\load$.
As we assume each element of the loadings are \emph{a priori} i.i.d.~$\ndist{0}{1}$, the gradient of the log-prior $\grad{\logdens{\load}}{\load}$ can be computed simply as $\pderiv{}{\loadij}\logdens{\load} = -\loadij$ for $\traitItoo$, $\facItoo$.

As computing $\grad{\logcdens{\obsdata}{\load, \errprec,\tree}}{\load}$ directly via Equation \ref{eq:naiveLikelihood} scales $\bigo{\ntaxa^3\ntraits^3}$ and is intractable for most problems, we use the highly structured nature of the phylogeny to compute this gradient in $\bigo{\ntaxa\ntraits\nfac^3}$.
We omit explicit dependence on the tree structure $\tree$ and the error precision $\errprec$ from the calculations below in favor of simpler notation.
We calculate the gradient of the likelihood with respect to each column of the loadings $\loadcol{\traitI}$ individually to accommodate variation in the missing data structure across traits.
\begin{equation}\label{eq:loadGrad}
\begin{aligned}
\grad{\log\cdens{\obsdata}{\load}}{\loadcol{\traitI}} &= \frac{1}{\cdens{\obsdata}{\load}}\grad{\cdens{\obsdata}{\load}}{\loadcol{\traitI}} \\
&= \frac{1}{\cdens{\obsdata}{\load}} \gradvar{\loadcol{\traitI}}\left[\int\cdens{\obsdata}{\fac, \load}\dens{\fac}\diff\fac\right] \\
&= \frac{1}{\cdens{\obsdata}{\load}} \int \dens{\fac} \grad{\cdens{\obsdata}{\fac, \load}}{\loadcol{\traitI}} \diff\fac .
\end{aligned}
\end{equation}
Based on the fact that $\cdist{\data}{\fac, \load} \sim \mndist{\fac\load}{\I{\ntaxa}}{\errprec\inv}$ and our ignorable missing measurements assumption, we have
\def\loadExp{\errprecel{\traitI} \left(\fac\trans\taxamismat{\traitI}\obsdatacol{\traitI} - \fac\trans\taxamismat{\traitI}\fac\loadcol{\traitI}\right)}
\begin{equation}
\begin{aligned}
\grad{\cdens{\obsdata}{\fac, \load}}{\loadcol{\traitI}} = \cdens{\obsdata}{\fac, \load} \loadExp,
\end{aligned}
\end{equation}
where $\obsdatacol{\traitI}$ is the $\traitI^{\text{th}}$ column of $\obsdata$ and $\taxamismat{\traitI} = \diagOp{\misind{1\traitI}\too\misind{\ntaxa\traitI}}$ is a diagonal matrix of observed-data indicators (see \siSec~\siref{ap:gradCalc} for detailed calculations).
Using this result in Equation \ref{eq:loadGrad}, we calculate
\begin{equation}
\begin{aligned}
\grad{\log\cdens{\obsdata}{\load}}{\loadcol{\traitI}} &= %
\int \frac{\dens{\fac} \cdens{\obsdata}{\fac, \load}}{\cdens{\obsdata}{\load}} \loadExp \diff\fac \\
&= \int \cdens{\fac}{\obsdata, \load} \loadExp \diff\fac \\
&= \cexpectOp{\loadExp}{\obsdata,\load} \\
&= \errprecel{\traitI}\cexpectOp{\fac\trans}{\obsdata, \load}\taxamismat{\traitI}\obsdatacol{\traitI} - \errprecel{\traitI}\cexpectOp{\fac\trans\taxamismat{\traitI}\fac}{\obsdata, \load} \loadcol{\traitI} .
\end{aligned}
\end{equation}
Note that
\begin{equation}
\begin{aligned}
\cexpectOp{\fac\trans\taxamismat{\traitI}\fac}{\obsdata, \load} &= \sum_{\taxaI = 1}^\ntaxa \misind{\taxaI\traitI}\cexpectOp{\facrow{\taxaI}\facrow{\taxaI}\trans}{\obsdata, \load} \\
&= \sum_{\taxaI = 1}^\ntaxa \misind{\taxaI\traitI}\cvarOp{\facrow{\taxaI}}{\obsdata, \load} + \misind{\taxaI\traitI}\cexpectOp{\facrow{\taxaI}}{\obsdata, \load}\cexpectOp{\facrow{\taxaI}}{\obsdata, \load}\trans.
\end{aligned}
\end{equation}
We compute $\cexpectOp{\facrow{\taxaI}}{\obsdata, \load}$ and $\cvarOp{\facrow{\taxaI}}{\obsdata, \load}$ for $\taxaI = 1\too\ntaxa$ in $\bigo{\ntaxa\ntraits\nfac^3}$ via the pre-order algorithms independently developed by \citet{bastide2018inference} and \citet{alex2019relaxed}.

\subsubsection{Orthogonality Constraint and Post-Processing} \label{sec:svdProcessing}
While both the Gibbs and HMC samplers above can enforce the structured sparsity constraint, neither can enforce the orthogonality constraint directly.
However, as both the likelihood and i.i.d.~prior are invariant with respect to orthonormal rotations of $\load$, applying such a rotation to all posterior samples via post-processing results in a valid posterior.
We can easily rotate the loadings to have orthogonal rows with descending norms via singular value decomposition (see \siSec~\siref{ap:svd} for details).

\subsection{Loadings Under the Orthogonal Shrinkage Prior}\label{sec:shrinkInference}
Both samplers above are incompatible with the orthogonal shrinkage prior from Section \ref{sec:shrink} as 1) they cannot enforce the orthogonality constraint directly and 2) post-processing is invalid because the prior is not rotationally invariant.
Therefore, we sample directly from the full conditional distributions of both $\loadScale$ and $\loadOrtho$ rather than their product $\load$.

\subsubsection{Geodesic HMC Sampler on the Orthonormal Component \texorpdfstring{$\loadOrtho$}{V}}\label{sec:hmcOrtho}

Requiring $\loadOrtho\trans \in \stiefel{\nfac}{\ntraits}$ allows us to employ existing techniques for sampling from the Stiefel manifold.
Geodesic HMC \citep{byrne2013geodesic} uses the same fundamental principles of standard HMC, but progresses parameters along geodesics on manifolds rather than through Euclidean space. %
This procedure also relies on the gradient of the log-posterior with respect to the parameter of interest.
As such, to efficiently employ geodesic HMC to update the orthonormal matrix $\loadOrtho$, we must efficiently compute the gradient
\begin{equation}
\grad{\logcdens{\loadOrtho}{\obsdata, \loadScale, \errprec, \tree}}{\loadOrtho} = \grad{\logcdens{\obsdata}{\loadOrtho, \loadScale, \errprec, \tree}}{\loadOrtho} +  \grad{\logdens{\loadOrtho}}{\loadOrtho}.
\end{equation}
As noted in Section \ref{sec:shrink}, we place a uniform prior on $\loadOrtho$ and can therefore ignore $\grad{\logdens{\loadOrtho}}{\loadOrtho}$.
Using our calculations for $\grad{\logcdens{\obsdata}{\load, \errprec, \tree}}{\load}$ from Section \ref{sec:hmcLoadings}, the chain rule provides a simple formula for the gradient of the likelihood with respect to $\loadOrtho$ as $\load = \loadScale\loadOrtho$:
\begin{equation}
\grad{\logcdens{\obsdata}{\loadOrtho,\loadScale,\errprec, \tree}}{\loadOrtho} = \loadScale \grad{\logcdens{\obsdata}{\load, \errprec, \tree}}{\load}.
\end{equation}
We then use this gradient in the geodesic HMC algorithm of \citet{holbrook2016bayesian} to sample from the full conditional distribution of $\loadOrtho$.

\subsubsection{Gibbs Sampler on the Diagonal Scale Component \texorpdfstring{$\loadScale$}{sigma}}
While we can employ HMC to sample from $\cdist{\loadScale}{\obsdata, \loadOrtho, \errprec, \tree}$, our implementation did not mix well in practice.
We develop a Gibbs sampler to draw from $\cdist{\loadScale}{\obsdata, \loadOrtho, \errprec, \fac}$ as an efficient alternative that relies on the data augmentation of $\fac$ in \siSec~\siref{ap:preOrder}.
We define the $\nfac$-vector $\loadScales$ such that $\loadScale = \diagOp{\loadScales}$ and sample $\loadScales$ as follows (see \siSec~\siref{sec:scaleGibbsDetails} for derivation):
\def\scalePrior{\logdens{\loadScales}}
\def\orthocoli{\loadOrthocol{\traitI}}
\def\errpreci{\errprecel{\traitI}}
\def\errTrait{\left(\fac\loadScale\orthocoli - \obsdatacol{\traitI}\right)}
\def\taxamismati{\taxamismat{\traitI}}
\def\scalePostPrecLong{\diagOp{\gshrinks} + \sumTraits\errpreci\diagOp{\orthocoli}\fac\trans\taxamismati\fac\diagOp{\orthocoli}}
\def\scalePostMeanSub{\sumTraits\errpreci\diagOp{\orthocoli}\fac\trans\taxamismati\obsdatacol{\traitI}}
\begin{equation}\label{eq:scaleGibbs}
\begin{aligned}
\cdist{\loadScales}{\obsdata,\fac,\loadOrtho,\errprec} &\sim \mvnDist{\scalePostMean}{\scalePostPrec\inv}\text{, where}\\
\scalePostPrec &= \scalePostPrecLong,\\
\scalePostMean &= \scalePostPrec\inv\left(\scalePostMeanSub\right),
\end{aligned}
\end{equation}
$\gshrinks = (\gshrinkel{1}\too\gshrinkel{\nfac})$ and $\orthocoli$ is the $\traitI^{\text{th}}$ column of $\loadOrtho$.

While the prior encourages the elements of $\loadScales$ to have descending absolute value, it does not enforce this constraint strictly.
As discussed in Section \ref{sec:shrink}, for some problems a strict ordering with forced spacing may be necessary in practice for full identifiability.
In these cases we employ a rejection sampler where we draw from the full conditional distribution of $\loadScales$ using the unrestricted multivariate normal distribution and reject any samples that do not conform to the particular constraint.
As the unconstrained prior already induces a soft ordering, we find that this rejection sampler typically has high acceptance probability.

\subsubsection{Gibbs Sampler on the Precision Multipliers}
We must also sample from the shrinkage multipliers $\shrinkMult{1}\too\shrinkMult{\nfac}$ when using the shrinkage prior on the loadings.
\citet[][Section 3.1, Step 5]{bhattacharya2011sparse} develop a conjugate Gibbs sampler for these multipliers that we apply directly to this model. %

\subsection{Sign Constraint on the Loadings}

\def\maxInd{\traitI_\mathrm{max}}
\def\sumStates{\sum_{\state = 1}^\nstates}
\def\absStateMean{\bar{\loadel{}}_{\traitI\facI}^\mathrm{abs}}
Regardless of which prior (i.i.d.~vs.~orthogonal shrinkage) or constraint (sparsity vs.~orthogonality) we choose, we must enforce a sign constraint on a single element in each row of $\load$ for full identifiability (see \siSec~\siref{ap:signConstraint} for details).

\subsection{Gibbs Sampler on the Error Precisions \texorpdfstring{$\errprec$}{Lambda}} \label{sec:errgibbs}

We sample from $\cdist{\errprec}{\fac, \obsdata, \load}$ using the same procedure as \citet{tolkoff2017phylogenetic} in conjunction with the data augmentation algorithm in \siSec~\siref{ap:preOrder} (see \siSec~\siref{ap:precInference} for details).

\section{Computational Efficiency}\label{sec:timing}
\def\nsims{3}
\def\nsimstext{three}

\begin{figure}
	\centering
	\includegraphics[width=0.6\linewidth]{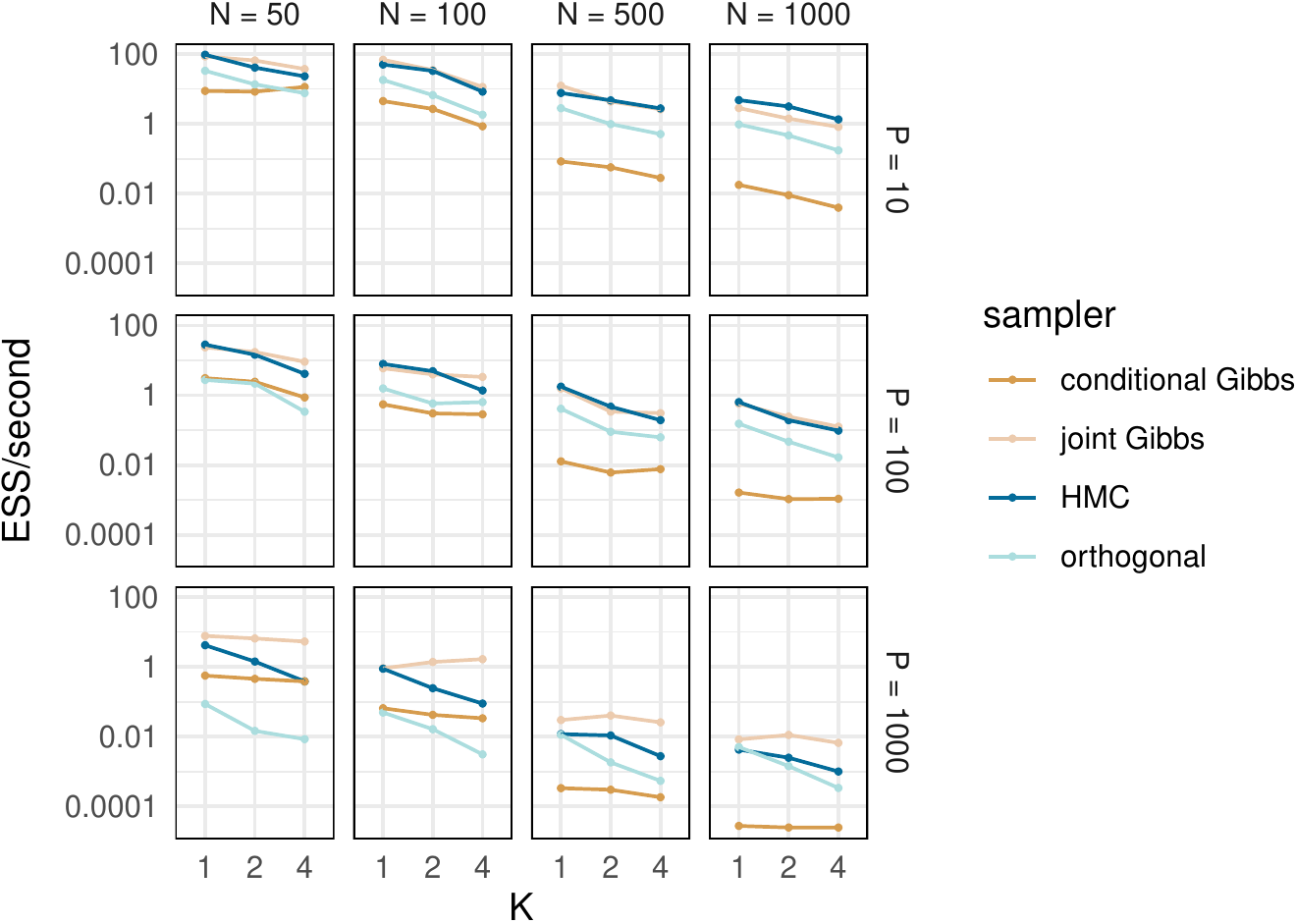}
	\caption{Timing comparison between inference regimes. We run three MCMC chain simulations for each combination of $\ntaxa$, $\ntraits$, $\nfac$ and sampler and present the average minimum ESS per second for each.
		The ``conditional Gibbs" sampler refers to the methods used by \citet{tolkoff2017phylogenetic}.
		The ``joint Gibbs", ``HMC" and ``orthogonal" samplers refer to the methods presented in Sections \ref{sec:gibbsLoadings}, \ref{sec:hmcLoadings} and \ref{sec:shrinkInference} respectively.
		Values are available in \siTab~\siref{tb:timing}.
	}
	\label{fig:timing}
\end{figure}

We compare the computational efficiency of the inference regimes discussed in Sections \ref{sec:gibbsLoadings}, \ref{sec:hmcLoadings} and \ref{sec:shrinkInference} with that of \citet{tolkoff2017phylogenetic}. %
To understand performance across a wide range of situations, we simulate \nsimstext~unique data sets for all 36 combinations of $\ntaxa\in\{50, 100, 500, 1000\}$, $\ntraits\in\{10, 100, 1000\}$ and $\nfac\in\{1, 2, 4\}$
(see \siSec~\siref{ap:timing} for simulation details).
To understand the relative performance of each inference regime, we compare the effective sample size (ESS) per second of the loadings across all four samplers (see \siSec~\siref{ap:ess} for details) and report our results in Figure \ref{fig:timing}.

Compared against the conditional Gibbs sampler of \citet{tolkoff2017phylogenetic}, both our joint Gibbs and HMC samplers under the i.i.d.~prior consistently yield efficiency gains of an order of magnitude in small-$\ntaxa$ data sets and two orders of magnitude in big-$\ntaxa$ data sets.
While the sampling regime under the orthogonal shrinkage prior is slower than either the joint Gibbs or HMC sampler (and even the conditional Gibbs sampler for small-$\ntaxa$, big-$\ntraits$), it has clear advantages over the others that we discuss in Section \ref{sec:priorChoice}.

\section{Principled Analysis Plan} \label{sec:analysisPlan}

The modeling decisions required for Bayesian factor analysis can be daunting.
In addition to the priors, identifiability constraints and sampling procedures discussed above, researchers must also choose an appropriate number of factors $\nfac$.
Making such choices in a principled manner is challenging, and experimenting with different combinations to determine which ``work best" is time consuming and opens the door to modeling decisions based on publication concerns.
We propose a generalizable analysis plan to guide researchers through this process.
To aid researchers seeking to employ phylogenetic factor analysis specifically, we also develop software tools that codify this plan and automate core procedures.

\subsection{Choosing the Loadings Constraint} \label{sec:constraint}
The decision to apply the sparsity constraint versus the orthogonality constraint depends on the biological question of interest.
While the sparsity constraint induces ordering onto the traits, this ordering can be desirable under certain circumstances.
For example, if one is trying to isolate the effects of a particular set of traits, placing those traits first in conjunction with the upper triangular constraint ensures that they will load only onto the first few factors and all subsequent factors will be independent of their influence. %
If one does not want to apply such an ordering, the orthogonality constraint may be a better alternative.
We emphasize, however, that the orthogonality constraint is no less restrictive than the sparsity constraint; rather, it replaces a series of potentially arbitrary modeling decisions (i.e.~the ordering of the first $\nfac$ traits) with a single, perhaps equally arbitrary, constraint.

Researchers can also apply a hybrid approach where one or more traits load only onto a certain factor(s) while the remaining traits are free to load onto all factors.
If the specific sparsity structure is not sufficient to induce identifiability, then any unconstrained sub-matrices of the loadings would require rotation to orthogonality.
We present a simple example of this in Section \ref{sec:mammals}, where the the first trait (body mass) loads only onto the first factor and the remaining traits load onto all $\nfac$ factors.
In this case, the first row of the loadings is identifiable and captures mass-dependent relationships, while the sub-matrix composed of rows $2\too\nfac$ and columns $2\too\traitI$ is rotated to orthogonality via post-processing.

\subsection{Choosing the Loadings Prior} \label{sec:priorChoice}
\begin{figure}[!htbp]
	\centering
	\includegraphics[width=\linewidth]{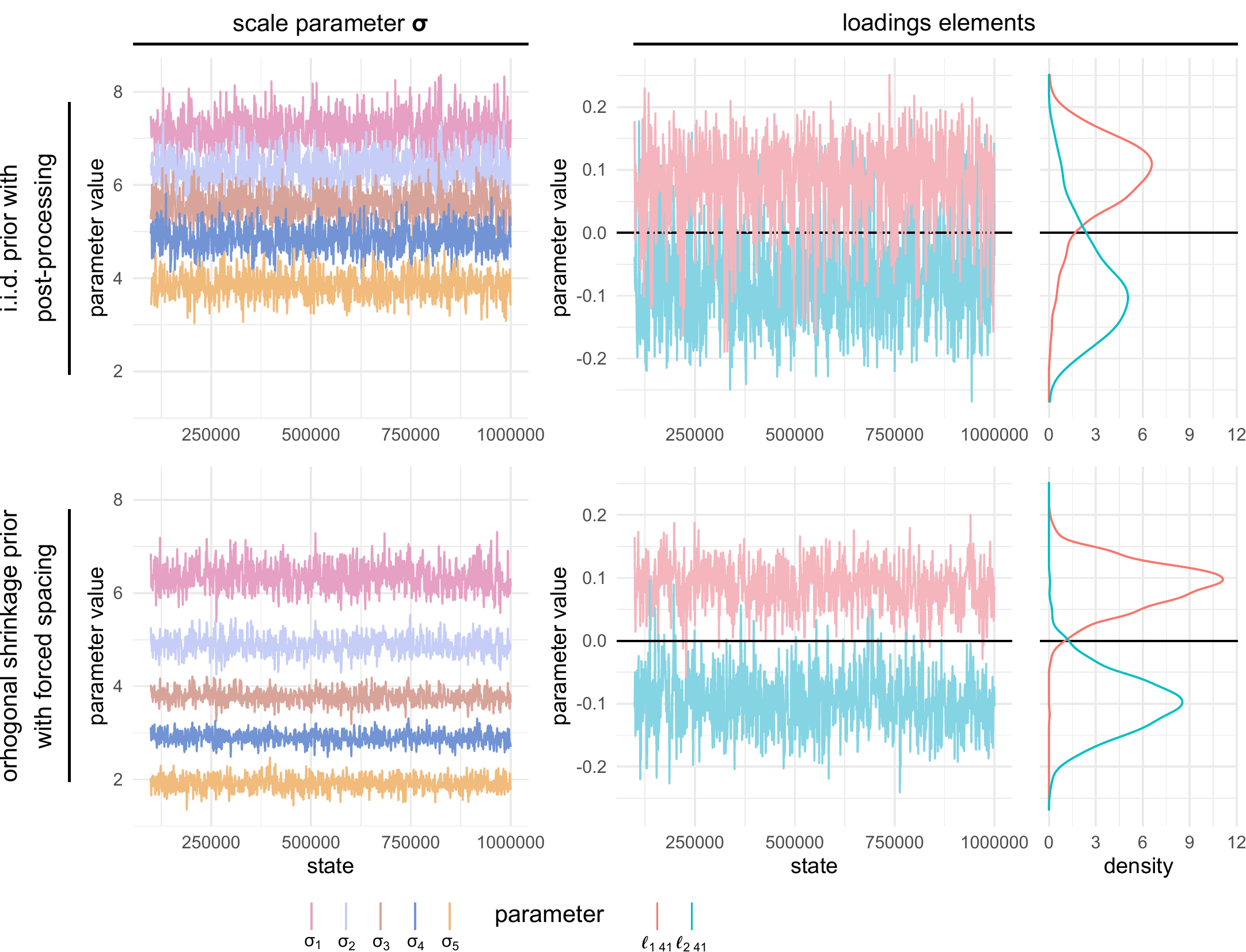}
	\caption{Trace plots of relevant parameters from analysis in Section \ref{sec:yeast}.
		Estimates under the i.i.d.~Gaussian prior are characteristic of poorly-identifiable conditions (the scales $\loadScales$ are overlapping resulting in label switching / row-wise convolution of the loadings).
		The shrinkage prior with forced spacing ($\scaleThreshold = 0.8$) largely eliminates this problem.}
	\label{fig:trace}
\end{figure}

Those choosing the sparsity (or hybrid) constraint must use the i.i.d.~prior on the loadings, as orthogonality is implicit in our definition of the shrinkage prior.
For those opting for the orthogonality constraint, we recommend choosing a prior based on the characteristics of the specific application.
For big-$\ntaxa$ data sets ($\ntaxa > 1000$) the geodesic HMC sampler on $\loadOrtho$ under the shrinkage prior may be prohibitively slow (particularly when combined with big-$\ntraits$), %
and we suggest using the i.i.d.~prior with post-processing.
One serious limitation of the post-processing regime, however, is the potential for label switching \citep{celeux1998bayesian} or row-wise convolution of the posterior of the loadings if the scale parameters $\loadScales$ have posterior means that are close to each other or large variances.
In this case, the trace plots of the scales $\loadScales$ will appear to be touching, and the posterior on the loadings themselves will often have heavy tails that overlap with 0 (see Figure \ref{fig:trace}).
Conversely, if the posteriors of the loadings are clearly distinct, then one can safely assume this phenomenon is not occurring and the post-processing regime is appropriate.
Conveniently it is in these big-$\ntaxa$ data sets where lack of identifiability of the loadings is at a minimum, and we find that the row-wise convolution phenomenon does not typically pose a problem under these circumstances.

For small or moderate $\ntaxa$ data sets, we still suggest attempting the i.i.d.~sampler with post-processing, but we caution users to look for evidence of label switching.
If such evidence exists, we recommend using the shrinkage prior with forced ordering and separation.

\subsection{Constraining the Number of Factors}
\def\maxFac{\nfac_{\mathrm{max}}}

We propose cross-validation for identifying the number of factors with optimal predictive performance.
In the case of the i.i.d.~prior, this procedure compares models with different number of factors directly, while in the case of the orthogonal shrinkage prior it tunes the strength of the shrinkage on the loadings scales.

We fully recognize that complex evolutionary processes do not, in reality, conform exactly to the phylogenetic latent factor model (or any tractable statistical model).
As such, we caution against seeking to identify the ``true" number of underlying evolutionary processes driving the phenotypes of interest, as such ground truth likely does not exist.
Rather, we encourage researchers to use this model selection procedure to identify the limitations of the information available in a particular data set and the model's ability to extract it.
For example, if model selection determines that a four factor model provides optimal predictive performance, one should be wary of interpreting results from a model with greater than four factors as it is likely some of the perceived signal is an artifact of noise in the data.

\subsubsection{Choosing \texorpdfstring{$\maxFac$}{Kmax}}

Prior to model selection, one must choose some maximum number of factors $\maxFac$ that balances model interpretability, flexibility, identifiability and tractability.
Models with more factors are inherently more flexible and can potentially capture more information about underlying biological phenomena.
However, interpretation becomes challenging as the number of factors increases. %
While the model with optimal predictive performance may have $\nfac < \maxFac$, one should be open to interpreting a model where $\nfac=\maxFac$.
Limiting $\maxFac$ provides additional benefits, as 1) the identifiability challenges discussed in Section \ref{sec:priorChoice} intensify with increasing $\nfac$ and 2) inference scales cubically with $\nfac$ and some big-$\nfac$ models may be intractable.
In practice, we settle on $\maxFac=5$ for most examples below, as we find that the computation time and identifiability issues are typically manageable at $\nfac = 5$ and feel most researchers would rarely need to interpret more than five factors.

\subsubsection{Model Selection} \label{sec:modelSelection}
Our model selection strategy seeks to identify the shrinkage strength (when using the shrinkage prior) or number of factors (when using the i.i.d.~prior) that provides optimal predictive performance via cross-validation.
To this end, we posit $\nmodels$ sub-models characterized by the meta-parameters $\priorparams{1}\too\priorparams{\nmodels}$.
Under the i.i.d.~prior, $\priorparamsi = \modparami{\nfac}$ is the number of factors in model $\modelI$.
For example, our default for the i.i.d.~prior assumes $\maxFac = 5$ and $\nmodels = 5$ models with $(\modparam{\nfac}{1}\too\modparam{\nfac}{\nmodels}) = (1, 2, 3, 4, 5)$.
Under the shrinkage prior, let $\priorparamsi = \{\modparami{\multshapes}, \modparami{\multrates}\}$ be the shapes and rates, respectively, of the gamma priors on the shrinkage multipliers $\shrinkMult{1}\too\shrinkMult{\nfac}$ for model $\modelI$.
We typically retain $\maxFac = 5$ and define the 5 sub-models as $\modparami{\multshapes} = 10^{(\modelI + 1)/2}\onevec{\maxFac}$ and $\modparami{\multrates} = \onevec{\maxFac}$ for $\modelI = 1\too5$.

We evaluate the predictive performance of each model on $\nrep$ replicate data sets via $\nrep$-fold cross-validation.
For each replicate $\repi = 1\too\nrep$, we randomly partition the observed data $\obsdata$ into a training set $\traindata$ containing ($100 - \frac{100}{\nrep}$)\% of the data and a validation set $\valdata$ with the remaining $\frac{100}{\nrep}$\% such that each observation occurs in exactly one validation set.

Let $\allparams = \{\load, \errprec\}$ be the model parameters relevant to the likelihood.
We first approximate $\cdens{\allparams}{\traindata, \priorparamsi}$ for $\modelItoo$, $\repi=1\too\nrep$ via MCMC simulation as described in Section \ref{sec:inference}.
We then compute the expected log predictive density \citep{gelman2013bayesian} $\lpdi = \expectOp{\logcdens{\valdata}{\traindata,\allparamsposti}}$ for $\modelItoo$, $\repi=1\too\nrep$, where $\allparamsposti$ is a random variable with density $\cdens{\allparams}{\traindata, \priorparamsi}$.
We select $\priorparams{\bestModel}$, where $\bestModel = \argmax_\modelI \frac{1}{\nrep}\sum_\repi \lpdi$, as the optimal model and approximate $\cdens{\load,\errprec}{\obsdata, \priorparams{\bestModel}}$ as the final step in the analysis plan.

\def\software{PhylogeneticFactorAnalysis.jl}
\subsection{Software Implementation}
We implement all inference procedures in Section \ref{sec:inference} in the Bayesian phylogenetic inference software BEAST \citep{BEAST}.
While BEAST is an extraordinarily flexible tool, this flexibility can result in a user experience that is overwhelming for the uninitiated.

We develop the Julia package \software~(see \href{https://gabehassler.github.io/PhylogeneticFactorAnalysis.jl/dev/}{documentation}) to both simplify the BEAST user experience (in the context of PFA) and automate model selection, post-processing, diagnostics and plotting.
Users must input the trait data, a phylogenetic tree, the identifiability constraint on the loadings and the prior on the loadings.
Users may also optionally specify other modeling decisions such as whether to standardize the trait data (which we recommend) and the model selection meta-parameters as well as a BEAST input file with instructions for inferring the phylogenetic tree from sequence data.

After receiving appropriate input, \software~automatically performs model selection and outputs a series of files including the sub-sampled MCMC realizations and plots of both the loadings (see Figures \ref{fig:aquiLoad}B, \ref{fig:yeastLoadings} and \ref{fig:mammalsPhylo}A) and factors on the tree (see Figures \ref{fig:mammalsPhylo}B and \ref{fig:nwm}B) using the ggplot2 \citep{ggplot2} and ggtree \citep{ggtree} plotting libraries.

\section{Example Analyses}\label{sec:results}

We demonstrate the utility of these methods in the four examples below.
Unless otherwise noted, all data are standardized on a per-trait basis (i.e.~subtracting the trait mean and dividing the by the trait standard deviation) prior to analysis.

\def\aquilegia{\emph{Aquilegia}}
\def\arbarg{x}
\def\disclevel{d}
\newcommand{\ndisc}[1]{D_{#1}}

\newcommand\hpdi[2]{(#1\%-#2\%)}
\newcommand\hpdiFirst[2]{(#1\%-#2\% HPD interval)}

\subsection{Pollinator-Flower Co-evolution in \aquilegia}
The intimate relationship between plants and their pollinators has played a defining role in the evolution of angiosperms \citep[see][]{kay2009role, van2012phylogenetic}.
Here we re-evaluate the relationship between floral phenotypes and pollinators in the genus \aquilegia~(columbines).
\citet{whittall2007} identify three primary \aquilegia~``pollination syndromes" associated with bumblebees, hummingbirds and hawk moths respectively.
\citet{tolkoff2017phylogenetic} apply phylogenetic factor analysis to study the relationship between 11 floral phenotypes and these pollination syndromes in \aquilegia~and identify two factors, only one of which is associated with pollinator type.  %

We re-evaluate this previous work for two reasons.
First, \citet{tolkoff2017phylogenetic} assume the upper-triangular constraint on the loadings which requires that the vertical angle of the flower loads only onto the first factor.
Additionally, we are eager to compare our cross-validation model selection procedure with the marginal likelihood-based approach of \citet{tolkoff2017phylogenetic} which identifies a two-factor model as having greatest posterior support.

\begin{figure}[!htbp]
	\centering
	\includegraphics[width=\linewidth]{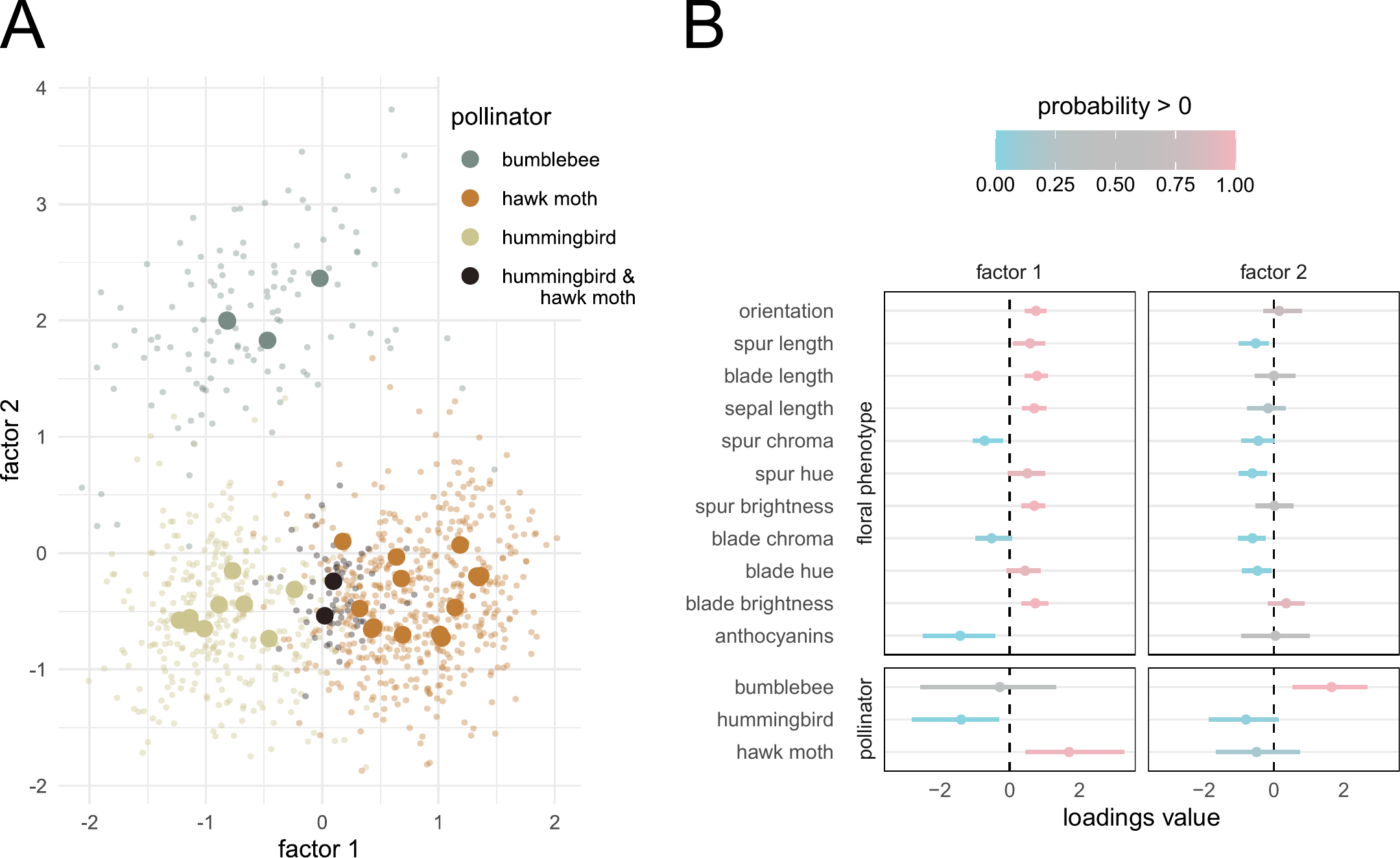}
	\caption{
		\aquilegia~results.
		\textbf{A)} Factor values colored by pollinator(s) for each species of \aquilegia.
		Large, solid points represent posterior means for each species.
		Small, transparent points represent a random sample from the posterior distribution of the factors.
		\textbf{B)} Posterior summary of the loadings matrix.
		Dots represent posterior means while bars cover the 95\% highest posterior density (HPD) interval.
		Colors represent the posterior probability that the parameter is greater than 0.
		While the second factor clearly separates the bumblebee-pollinated plants from the others, the first factor captures a more gradual transition from hummingbird pollination to hawk moth pollination.
		}
	\label{fig:aquiLoad}
\end{figure}

As four of the traits (anthocyanin production and the three pollination syndromes) are binary, we follow \citet{tolkoff2017phylogenetic} in adapting the latent-liability model of \citet{cybis2015assessing} to the latent factor model (see \siSec~\siref{ap:latentLiability}).
We use the i.i.d.~prior with orthogonality constraint, and our model selection procedure, indeed, identifies two factors.
We present our results in Figure \ref{fig:aquiLoad}. %
The first factor captures patterns differentiating hummingbird-pollinated plants from hawk moth-pollinated plants,
while the second factor appears to separate the bumblebee pollinated flowers from the other two pollination syndromes. %
Note that in Figure \ref{fig:aquiLoad}A, the first factor falls along a relatively uniform continuum, while the second factor has a clear out-group consisting of the bumblebee-pollinated plants.
While only two taxa are coded as being pollinated by both hummingbirds and hawk moths, this suggests that non-bumblebee \aquilegia~pollination strategies may lie on a continuum rather than strict a hawk moth/hummingbird dichotomy, and it is possible that many of the plants listed as having a single pollinator in reality attract both hummingbirds and hawk moths.

\subsection{Yeast Domestication} \label{sec:yeast}

\def\yeast{\textit{S.~cerevisiae}}
The brewer's yeast \emph{Saccharomyces cerevisiae} is essential to a variety of industrial applications due to its ability to convert sugars into ethanol, carbon dioxide and aroma compounds.
In addition to its well-known role in the production of fermented food and beverages, it also plays a key role in the production of of bio-fuels and serves as model organism for basic biological research.
Industrial strains within this species adapted to thrive within specialized environments and can withstand stress conditions often suited to the specific industrial niche they evolved in, such as ethanol, osmotic, acidic and temperature stresses.

Recent work by \citet{gallone2016domestication} and \citet{gallone2019interspecific} uses phylogenetic methods to study the domestication of \yeast~within industrial environments.
To elucidate the effects of domestication on yeast phenotypes, \citet{gallone2016domestication} sequence and phenotype 154 strains of industrial and wild \yeast.
The 82 phenotypes include numerous measurements of growth rates under varying environmental and nutrient stresses, the levels of production of various metabolites and the ability to reproduce sexually.

Domestication in plants and animals is typically characterized by limited reproduction outside of domestic contexts, increased yield and decreased tolerance to rare or novel environmental stressors \citep{doebley2006molecular, larson2014evolution}.
\citet{gallone2016domestication} observe these same patterns in the yeast strains they study, with additional niche-specific patterns of covariation.
While their analysis examines the specific hypotheses above, they do not employ a data-generative model of phenotypic evolution capable of studying broad changes across all measured phenotypes.

The phylogenetic latent factor model, however, is ideally suited for such a task.
We first infer a phylogenetic tree for the 154 phenotyped strains using the 2.8 megabase DNA sequence alignment of \citet{gallone2016domestication} (see \siSec~\siref{ap:yeastPhylo}). %
We fix this tree during model selection due to the computational costs of inferring the phylogeny.
Based on the principles discussed in Section \ref{sec:analysisPlan}, we opt for the orthogonality constraint, the orthogonal shrinkage prior with forced spacing ($\scaleThreshold=0.8$) and $\maxFac = 5$.
Our model selection procedure yields a final model with five significant factors.
For the final analysis we infer the tree jointly with factor model parameters using the same tree model in \siSec~\siref{ap:yeastPhylo}.
As the number of significant factors $\nfac$ is equal to the maximum $\maxFac$, we are confident any signal is biologically relevant but recognize we have not completely captured the full phenotypic covariance structure. %
That being said, the final factor captures only 7\% \hpdiFirst{5}{9} of the heritable variance and 3\% \hpdi{2}{4} of the total variance, suggesting that adding additional factors will yield diminishing returns at the expense of exacerbating identifiability challenges.

\begin{figure}[!htbp]
	\begin{center}
		\includegraphics[width=0.9\linewidth]{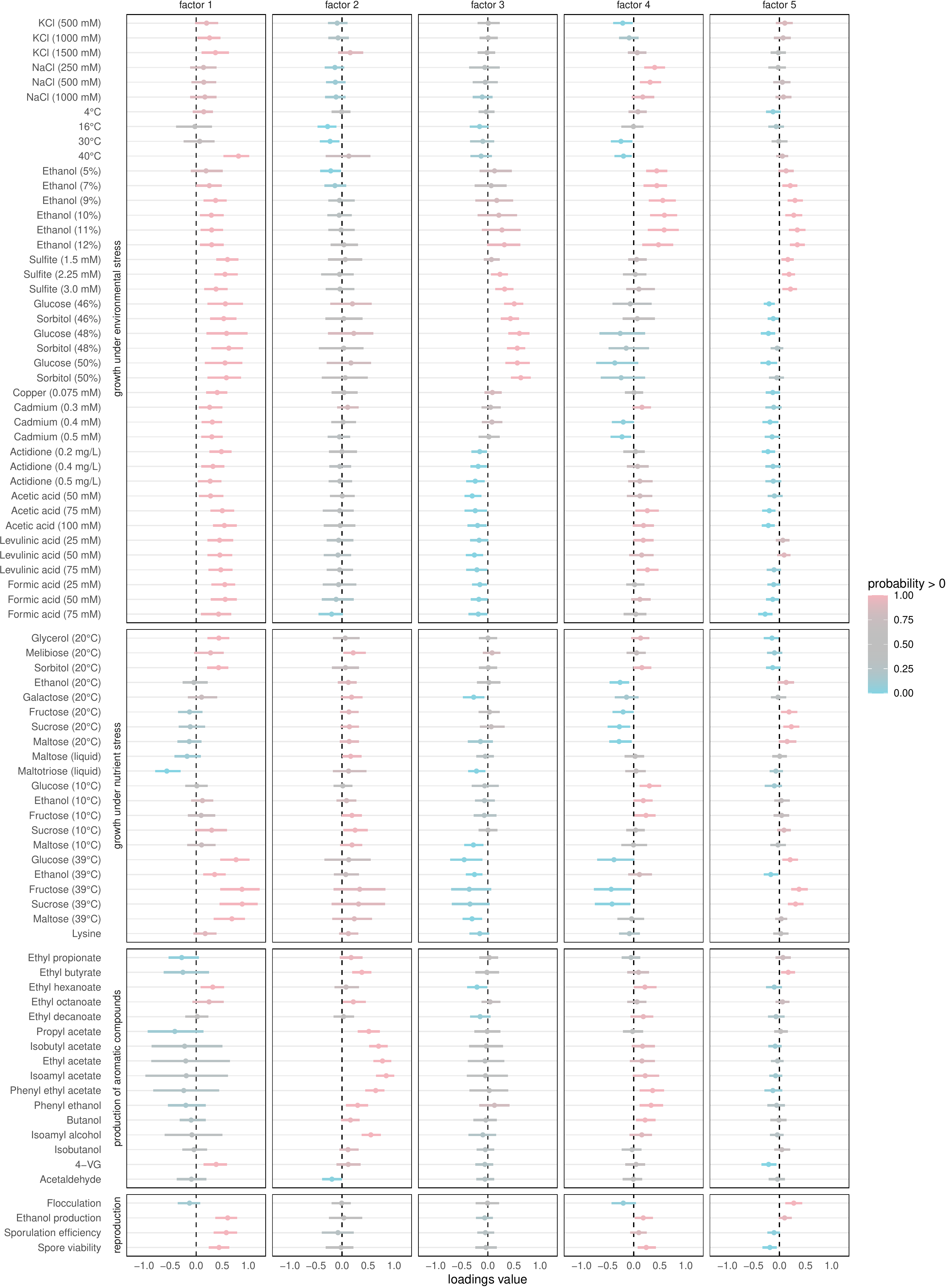}
		\caption{
			Posterior summary of loadings of 5-factor PFA on yeast data set.
			The first factor primarily captures differences associated with tolerance to environment and nutrient stress as well as reproductive ability.
			See Figure \ref{fig:aquiLoad}B for description of plot elements.
		}
		\label{fig:yeastLoadings}
	\end{center}
\end{figure}

\begin{figure}[!htbp]
	\begin{center}
		\includegraphics[width=0.985\linewidth]{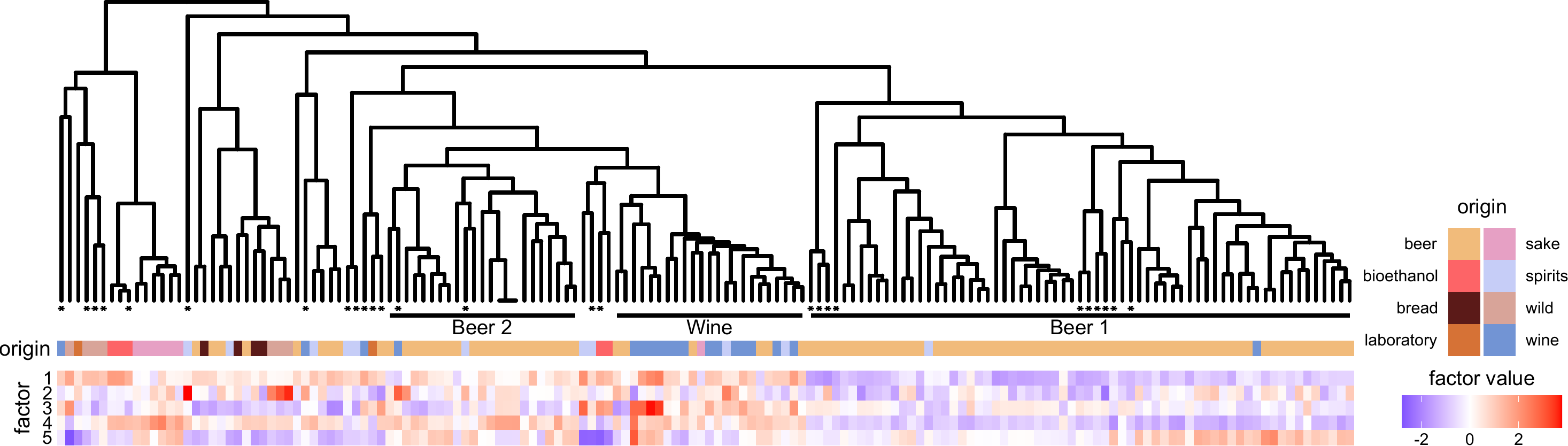}
		\caption{
			All five factors plotted on yeast phylogeny with strain origin.
			Stars at the tips indicate mosaic strains as identified by \citet{gallone2016domestication}.
			The first factor separates the Beer 1 clade from the remaining strains.
		}
		\label{fig:yeastAllPhylo}
	\end{center}
\end{figure}

We plot the loadings matrix in Figure \ref{fig:yeastLoadings} and factors along the phylogeny in Figure \ref{fig:yeastAllPhylo}.
For the first factor that accounts for 44\% \hpdi{33}{52} of the heritable variance, we observe a clear separation between strains in the Beer 1 clade and strains isolated from other fermentation processes and from the wild.
Notably, the domestication of beer strains in this clade led to an impaired sexual cycle as observed in the reduced sporulation efficiency and spore viability.
This loss of a functional %
sexual cycle is paired with the additional loss of tolerance to environment and nutrient stresses generally.
These stresses are not encountered during continuous growth in the nutrient-rich wort medium.
The higher tolerance to high temperature outside of Beer 1 might reflect other more cryptic specializations of non-Beer clade 1 strains selected for different industrial processes (e.g.~bioethanol or cocoa fermentation).
Beyond these general patterns, we also note specific traits selected for in the Beer 1 clade.
For example: strains within this clade do not produce 4-vinyl guaiacol (4-VG), a renown off-flavor in beer that is less relevant to other industrial niches.
Additionally, the first factor in this clade is associated with efficient utilization of maltotriose, an important carbon source in beer wort but rarely found in high concentrations in natural environments.
These results overall recapitulate one of the main findings of \citet{gallone2016domestication}: the transition from complex and variable natural niches to the stable, nutrient-rich, beer medium favored certain adaptations (e.g. efficient utilization of maltotriose) and accentuation of certain traits (lost of beer off-flavours) at the cost of becoming sub-optimal for survival in the wild.

We emphasize that in this dataset there are different domestication trajectories targeted to very diverse industrial processes, and the life histories of the different clades took separate paths that the additional factors likely capture. %

\def\nMammals{3,691}
\subsection{Mammalian Life History} \label{sec:mammals}

Life history strategies vary greatly across the tree of life.
Generally speaking, organisms exist along a spectrum between fast-reproducing species that produce many offspring with little investment into any single child and slow-reproducing species that invest relatively great time and energy into each of their (comparatively fewer) offspring \citep{pianka1970r}.
While allometric (size-dependent) constraints clearly influence these life history strategies \citep{boukal2014life}, pace-of-life theory predicts size-independent life-history variation as a major driver of phenotypic covariation \citep{Reynolds2003, Reale2010}.
Much work has been done evaluating these hypotheses across numerous taxonomic groups \citep[see][]{Blackburn1991, Bielby2007, Salguero2017}, but most studies are limited by methodologies that require complete data and scale poorly to very large trees and many traits.

We explore the evolution of mammalian life history using the PanTHERIA ecological database \citep{Jones2009}.
We select a sub-set of this data including body mass and 10 life history traits for the \nMammals~species with at least one non-missing observation.
While \citet{hassler2020inferring} explore a similar subset of the PanTHERIA data using a multivariate Brownian diffusion (MBD) model, the MBD model cannot partition the covariance structure into size-dependent and size-independent components.

\begin{figure}[!htbp]
	\begin{center}
		\includegraphics[width=0.9\linewidth]{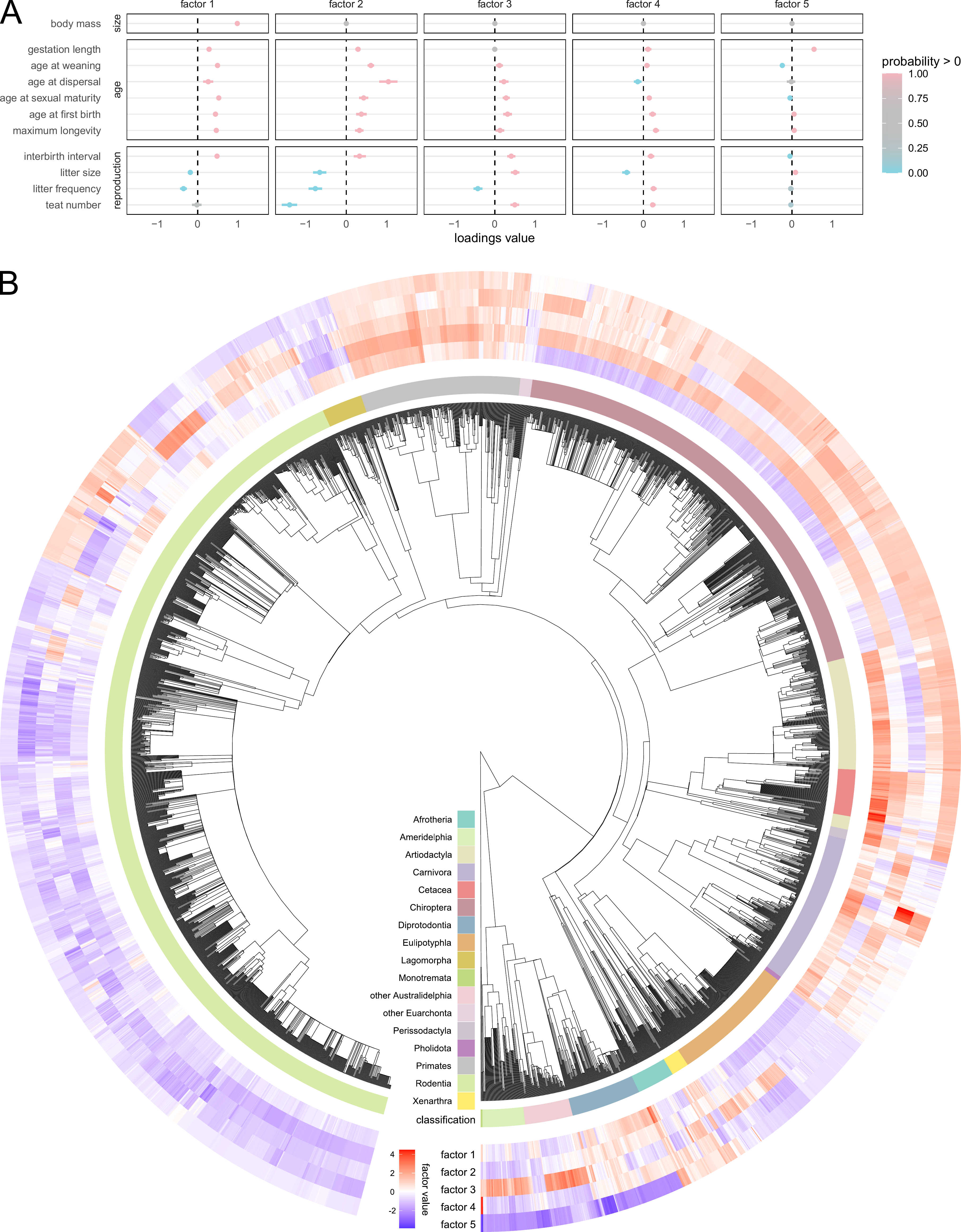}
	\end{center}
	\caption{
		Mammalian life history results. \textbf{A)} Posterior summary of loadings. Loadings of body size onto factors 2-5 is set to 0 \emph{a priori}.
		See Figure \ref{fig:aquiLoad}B for detailed description of figure elements.
		The first factor captures allometric relationships (by design), while the remaining factors capture size-independent relationships.
		\textbf{B)}	Evolution of factors along the mammalian phylogeny.
		Most factors appear to be strongly phylogenetically conserved throughout the tree, with large clades sharing similar factor values.
	}
	\label{fig:mammalsPhylo}
\end{figure}

PFA, however, is ideally suited to this task as we can structure the loadings matrix \emph{a priori} to reveal these relationships.
Specifically, we apply the hybrid constraint introduced in Section \ref{sec:constraint} where elements $\loadel{21}\too\loadel{\nfac1}$ are fixed to zero, forcing body mass to load only onto the first factor.
To avoid ordering the other life-history traits, we assume that the sub-matrix consisting of rows $2\too\nfac$ and columns $2\too\ntraits$ is orthogonal (which we enforce via post-processing).
We use the fixed tree of \citet{fritz2009}, which we prune to include only the \nMammals~taxa for which we have trait data.
We perform model selection assuming $\maxFac = 5$, with the optimal model having $\nfac=5$.
However, the first three factors explain 85\% of the heritable variance (with the last factor explaining only 4\%), suggesting that $\nfac=5$ is sufficient to capture the major patterns of variation in mammalian life-history evolution. %
We plot our results in Figure $\ref{fig:mammalsPhylo}$.

Consistent with the \citet{hassler2020inferring} analysis, body size is clearly associated with the ``slow" life history strategy (i.e.~smaller and less frequent litters, longer lives).
Notably, this allometric factor is not the dominant factor and explains only 16\% \hpdi{14}{18} of the heritable variance.
The second factor, however, explains 46\% \hpdi{42}{51} of this variance and clearly captures a size-independent fast-slow life history axis,
suggesting that size-independent life-history strategies play a major role in mammalian evolution.
As evident in Figure \ref{fig:mammalsPhylo}, this primary life-history axis (factor 2) varies independently of the allometric one (factor 1)
with examples of large/slow (cetaceans), large/fast (lagomorphs), small/slow (bats) and small/fast (rodents) taxonomic groups.
This primary life-history factor is well-conserved across the phylogenetic tree, with large taxonomic groups sharing life-history strategies.

Factors 3, 4 and 5 explain comparatively less of the heritable variance (23\%, 11\% and 4\% respectively).
Factors 3 and 4 appear to capture trade-offs between litter size and litter frequency, while the 5$^{\text{th}}$ factor primarily captures a negative relationship between weaning age and gestation length and is strongly expressed in monotremes and marsupials that employ different reproductive strategies than placental mammals.

\subsection{New World Monkey Cranial Morphology}\label{sec:monkeys}

While much effort has been devoted to studying the evolution of primate brain size, relatively few studies have focused on understanding diversity in brain morphology or shape.
Notable exceptions to this trend include \citet{aristide2016brain} and \citet{sansalone2020variation}.
Here we re-analyze the data presented in \citet{aristide2016brain}, that consist of 399 endocranial landmarks in 3-dimensional Euclidean space (standardized by generalized Procrustes analysis) for 48 species of New World monkey (NWM).
While \citet{aristide2016brain} perform principal component analysis on the Procrustes coordinates and use the principal component scores as traits in a larger evolutionary analysis, this procedure lacks a complete data-generative statistical model that explicitly accounts for uncertainty or noise in the shape data.

We simultaneously infer the phylogeny with the PFA parameters using DNA sequence alignments from \citet{aristide2015modeling} (see \siSec~\siref{ap:nwmTree} for details).
Preliminary results suggest 1) optimal predictive performance requires a very large number of factors ($>20$), which is unsurprising given the complexity of this data set, and 2) identifiability poses an unusually great challenge due to the ``small-$\ntaxa$ big-$\ntraits$" nature of the data.
As such, we settle on a 3-factor model with orthogonal shrinkage prior and strong shrinkage to maximize identifiability.
To maintain differences in scale between traits, we do not re-scale on a per-trait basis but rather divide all traits by the maximum per-trait standard deviation.

\begin{figure}[!ht]
	\centering
	\includegraphics[width=0.881\linewidth]{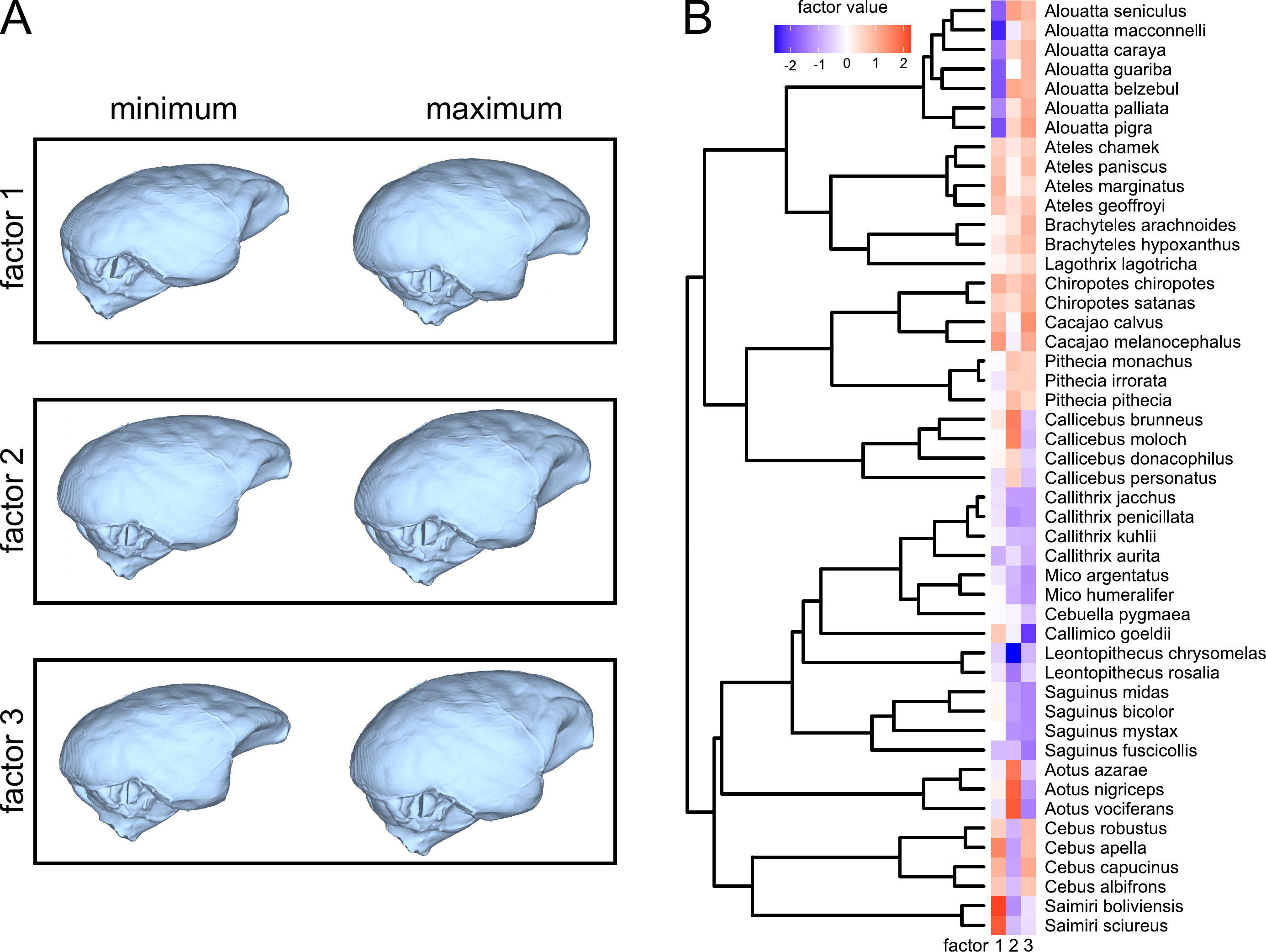}
	\caption{
		\textbf{A)} Influence of each factor on New World monkey brain shape.
		The first factor, capturing 60\% (57\%, 65\%) of the heritable variance, is highly correlated with the first principal component identified in \citet{aristide2016brain}. %
		\textbf{B)} Brain shape factors plotted along New World monkey phylogeny.
	}
	\label{fig:nwm}
\end{figure}

We plot the influence of each factor on brain shape and the evolution of these factors on the tree in Figure \ref{fig:nwm}.
These three factors capture similar patterns of variation as the first three principal components in \citet{aristide2016brain}, who identify %
several ecological processes associated with the evolution of these principal components.
As the latent factor model can capture uncertainty that PCA cannot, we are eager to re-evaluate these relationships via a more structured latent factor model that directly models the relationship between the brain shape factors and ecological phenotypes such as social structure or diet.
While preliminary results suggest that the first factor is correlated with relative brain volume (i.e.~brain volume divided by body mass) and social group size and that the third factor is correlated with body mass and absolute brain volume, we leave this more structured analysis as future work. %

\section{Discussion}\label{sec:discussion}

We develop a practical and scalable analysis plan requiring minimal user decisions enabled by computationally innovative inference procedures.
Previously, researchers performing phylogenetic factor analysis were limited by computational constraints and had to determine \emph{a priori} the ordering of the traits and optimal number of factors.
These computational and modeling advances are not independent but rather complement each other.
Our default model selection procedure requires 26 individual MCMC chain simulations (5-fold cross validation with 5 sets of meta-parameters plus the final run).
Such an analysis would be intractable for all but the smallest data sets using existing inference techniques.
However, our new inference procedures take only a few hours to run all 26 simulations for even the largest data sets we analyze.
Additionally, we have made these tools both flexible and accessible with the Julia package \href{https://github.com/gabehassler/PhylogeneticFactorAnalysis.jl}{\software},
which assembles and runs all BEAST input files, automatically performs model selection, plots the results and performs basic quality control.
Our implementation allows researchers to focus on big-picture modeling decisions and leave low-level implementation details to the software.

Limitations of this work that we plan to address in the future include the following.
First, while we can accommodate discrete phenotypes through the latent probit model of \citet{cybis2015assessing} (see \siSec~\siref{ap:latentLiability}), we notice both in our analysis and \citet{tolkoff2017phylogenetic} that the discrete parameters tend to have a far higher influence than their continuous counterparts (i.e.~the loadings entries associated with the discrete traits have greater magnitude than those associated with continuous traits).
This is likely due to the fact that we control the variance of the latent liabilities indirectly by fixing the discrete trait precisions $\errprec$ to a constant as do \citet{tolkoff2017phylogenetic}.
It is possible that the (potentially) inflated significance of these discrete traits can influence the loadings structure in unexpected ways, and we seek an alternative solution that places the continuous and discrete traits on more equal footing.

Second, there may be cases where the row-wise convolution of the loadings persists despite our efforts to induce identifiability.
Additional post-processing procedures developed for Bayesian mixture models \citep{rodriguez2014label} or multidimensional scaling \citep{okada2018post} may serve as solutions to these unusually convolved posteriors.
While preliminary work suggests that these methods can efficiently identify and deconvolve individual modes of multi-modal posteriors, we are concerned about their potential to identify non-existent signal in the data and believe a careful analysis of their properties is warranted.

Additionally, as proposed in Section \ref{sec:monkeys}, this work can be readily extended to incorporate parallel evolutionary models for different suites of traits.
In this framework, we could simultaneously perform factor analysis on a high-dimensional trait (e.g.~brain shape) and infer the evolutionary correlation between the latent factors and other phenotypes of interest (e.g.~brain size, diet, group size) using an MBD model.
Note that we could study relationships between multiple, distinct high-dimensional phenotypes as well from structural equation modeling paradigm \citep{lee2012basic}. %
While likelihood calculations under such models are straightforward given this and previous work, inferring the joint evolutionary covariance matrix requires additional inference machinery that we leave as future work.

Moreover, while we focus on the multivariate Brownian diffusion model of phenotypic evolution for simplicity, all inference machinery can be readily adapted to other Gaussian processes, such as the multivariate Ornstein–Uhlenbeck (OU) process \citep{hansen1997stabilizing}.
Indeed, the OU model and inference procedure of \citet{bastide2018inference} has already been implemented in BEAST and are easily integrated with the methods presented in this paper.

Finally, this work has applications to latent factor models beyond phylogenetics.
The gradients in Section \ref{sec:hmcLoadings} in particular are useful for hierarchical latent factor models generally.
Notably, they rely only on the conditional expectations of the latent factors rather than the values of the factors themselves, avoiding the need for potentially costly data augmentation when inferring the loadings.
While we take advantage of a Gaussian model of factor evolution to compute these expectations efficiently, the gradient calculations do not rely on any specific model and apply generally for all distributions with defined expectation and variance.
Additionally, the HMC procedure relying on the gradient can be readily adapted to non-standard priors or constraints in ways that conjugate Gibbs sampling cannot, which we demonstrate with our sampler over the Stiefel manifold.

\subsection*{Acknowledgments}
This work was supported through National Institutes of Health grants F31AI154824, K25AI153816, R01AI153044 and T32HG002536.
PL acknowledges funding from the European Research Council under the European Union's Horizon 2020 research and innovation programme (grant agreement no. 725422-ReservoirDOCS).
GB acknowledges support from the Interne Fondsen KU Leuven/Internal Funds KU Leuven under grant agreement C14/18/094, and from the Research Foundation - Flanders (“Fonds voor Wetenschappelijk Onderzoek - Vlaanderen,” G0E1420N, G098321N).
The Artic Network receives funding from the Wellcome Trust through project 206298/Z/17/Z.

\subsection*{Supplementary Material}
\begin{description}
	\item[Supplemental Information:] \siSec s 1 through 10 (pdf file)
	\item[Data and Code:] GitHub repository with data and code necessary for reproducing our analyses (\url{https://github.com/suchard-group/PhylogeneticFactorAnalysis})
	\item[PhylogenticFactorAnalysis.jl:] GitHub repository with Julia package for automating PFA analyses (\url{https://github.com/gabehassler/PhylogeneticFactorAnalysis.jl})
\end{description}

\bibliographystyle{chicago}
\bibliography{bibliography}

\clearpage

\setcounter{section}{0}
\setcounter{table}{0}
\setcounter{equation}{0}
\ifsiappended\else\setcounter{page}{1}\resetlinenumber\fi

\begin{center}
	\huge\bfseries Supplemental Information
\end{center}

\section{Post-order Traversal Likelihood Calculations} \label{ap:likeDetails}

\subsection{Partial Likelihood Calculations Under the Latent Factor Model} \label{ap:partials}

We present the detailed calculations from Equation \siref{eq:factorPartials}.
\begin{equation}\label{eq:partialDetails}
\begin{aligned}
\log\cdens{\obsdatarow{\taxaI}}{\facrow{\taxaI}, \load, \errprec} &= -\frac{\rank{\misprec{\taxaI}}}{2}\log2\pi + \frac{1}{2}\log\sdet{\misprec{\taxaI}}\\
&\hspace{0.5in} - \frac{1}{2}\left(\obsdatarow{\taxaI} - \load\trans\facrow{\taxaI}\right)\trans\misprec{\taxaI} \left(\obsdatarow{\taxaI} - \load\trans\facrow{\taxaI}\right) \\
&= -\frac{\ntraitobs{\taxaI}}{2}\log2\pi + \frac{1}{2}\sum_{\traitI = 1}^{\ntraits} \misind{\taxaI\traitI}\log \errprecel{\traitI}\\
&\hspace{.5in} - \frac{1}{2}\left[\facrow{\taxaI}\trans\load\trans\misprec{\taxaI}\load\trans\facrow{\taxaI} - 2\facrow{\taxaI}\trans\load\trans\misprec{\taxaI}\obsdatarow{\taxaI} + \obsdatarow{\taxaI}\trans\misprec{\taxaI}\obsdatarow{\taxaI}\right] \\
&= -\frac{\ntraitobs{\taxaI}}{2}\log2\pi + \frac{1}{2}\sum_{\traitI = 1}^{\ntraits} \misind{\taxaI\traitI}\log \errprecel{\traitI}  \\
&\hspace{.5in} - \frac{1}{2}\left[\left(\facrow{\taxaI} - \pmean{\taxaI}\right)\trans \pprec{\taxaI} \left(\facrow{\taxaI} - \pmean{\taxaI}\right)\right] - \frac{1}{2}\left[\obsdatarow{\taxaI}\trans\misprec{\taxaI}\obsdatarow{\taxaI}  - \pmean{\taxaI}\trans\pprec{\taxaI}\pmean{\taxaI}\right] \\
&= \log\prem{\taxaI} - \frac{\rank{\pprec{\taxaI}}}{2}\log 2\pi + \frac{1}{2}\log\sdet{\pprec{\taxaI}}\\
&\hspace{0.5in} - \frac{1}{2}\left[\left(\facrow{\taxaI} - \pmean{\taxaI}\right)\trans \pprec{\taxaI} \left(\facrow{\taxaI} - \pmean{\taxaI}\right)\right] \\
&= \log\prem{\taxaI} + \log\sndens{\facrow{\taxaI}}{\pmean{\taxaI}}{\pprec{\taxaI}},\text{ where }
\end{aligned}
\end{equation}
the partial precision $\pprec{\taxaI} = \load\misprec{\taxaI}\load\trans$, the partial mean $\pmean{\taxaI}$ is a (not necessarily unique) solution to $\pprec{\taxaI}\pmean{\taxaI} = \load\trans\misprec{\taxaI}\obsdatarow{\taxaI}$ and the remainder
\begin{equation}
\begin{aligned}
\log\prem{\taxaI} = -\frac{\ntraitobs{\taxaI} - \rank{\pprec{\taxaI}}}{2}&\log 2\pi + \frac{1}{2}\left(\sum_{\traitI = 1}^{\ntraits} \misind{\taxaI\traitI}\log \errprecel{\traitI} - \log\sdet{\pprec{\taxaI}}\right)\\
&- \frac{1}{2}\left[\obsdatarow{\taxaI}\trans\misprec{\taxaI}\obsdatarow{\taxaI}  - \pmean{\taxaI}\trans\pprec{\taxaI}\pmean{\taxaI}\right].
\end{aligned}
\end{equation}

\subsection{Special Inverse Calculations}\label{ap:inverse}

One challenge that the PFA model poses to this approach is that the partial precisions at the tips $\pprec{\taxaI}$ for $\taxaI=1,\hdots,\ntaxa$ may not be invertible via the pseudoinverse used by \citetsi{hassler2020inferringsup}.
The post-order traversal algorithm requires that for each internal node $\node{\nodeI}$ for $\nodeI = \ntaxa + 1,\hdots,2\ntaxa - 1$ in $\tree$, we must compute $\dpprec{\nodeI}$ such that $\cdens{\databelow{\nodeI}}{\facrow{\parent{\nodeI}}} = \prem{\nodeI}\sndens{\facrow{\parent{\nodeI}}}{\pmean{\nodeI}}{\dpprec{\nodeI}}$, where $\databelow{\nodeI}$ represents the trait values of all terminal descendants of node $\node{\nodeI}$.
In the PFA model, this  results in $\dpprec{\nodeI} = \left(\pprec{\nodeI}\inv + \blen{\nodeI}\I{\nfac} \right)\inv$.
However, it is possible that the initial partial precisions $\pprec{\taxaI}$ at the tip nodes $\node{1}, \hdots, \node{\ntaxa}$ may be rank-deficient.
This situation arises, for example, when the number of non-missing traits $\ntraitobs{\taxaI}$ at taxon $\taxaI$ is less than the number of factors $\nfac$.
To avoid this inversion, we use an algebraic slight-of-hand to compute $\dpprec{\nodeI}$ in terms of $\pprec{\nodeI}$ directly (rather than its non-existing inverse).
Specifically we use an identity for the inverse of the sum of two square matrices that has been discovered and forgotten several times \citepsi[see, for example, ][]{henderson1959estimationsup, henderson1981onsup}
\begin{equation}
\left(\matA + \matB\right)\inv = \matA\inv - \matA\inv\left(\I{} + \matB\matA\inv\right)\inv\matB\matA\inv.
\end{equation}
Applying this to our equation for $\dpprec{\nodeI}$, we get
\begin{equation}
\dpprec{\nodeI} = \pprec{\nodeI} - \blen{\nodeI}\pprec{\nodeI}\left(\I{k} + \blen{\nodeI}\pprec{\nodeI}\right)\inv\pprec{\nodeI}.
\end{equation}
Note that the matrix $\I{\nfac} + \blen{\nodeI}\pprec{\nodeI}$ is the sum of the positive semi-definite matrix $\blen{\nodeI}\pprec{\nodeI}$ with the positive definite matrix $\I{\nfac}$ and is therefore invertible.
As such, computing $\dpprec{\nodeI}$ is indeed possible and the \citetsi{hassler2020inferringsup} algorithm can proceed to compute the likelihood.

\section{Sampling from the Loadings \texorpdfstring{$\load$}{L} via Data Augmentation}

To employ the Gibbs sampler of \citetsi{tolkoff2017phylogeneticsup} to sample from the loading $\load$, we follow the procedure below:
\begin{enumerate}
	\item Sample from $\cdist{\fac}{\obsdata,\load,\errprec,\tree}$ via the pre-order algorithm of \citetsi{hassler2020inferringsup}
	\item Sample from $\cdist{\load}{\obsdata,\fac,\errprec}$ via the methods discussed in \citetsi{lopes2004bayesiansup}
\end{enumerate}

\subsection{Pre-Order Data Augmentation Algorithm}\label{ap:preOrder}

We seek to sample from $\cdist{\fac}{\obsdata,\load,\errprec,\tree}$ via the pre-order algorithm of \citetsi{hassler2020inferringsup}.
This procedure relies on first computing the statistics $\pmean{\taxaI}$ and $\pprec{\taxaI}$ such that
\begin{equation}
\cdens{\obsdatabelow{\taxaI}}{\facrow{\taxaI}, \load, \errprec, \tree} \propto \sndens{\facrow{\taxaI}}{\pmean{\taxaI}}{\pprec{\taxaI}}
\end{equation}
for $\taxaI = 1\too2\ntaxa - 1$ (i.e.~all nodes in the tree), where $\obsdatabelow{\taxaI}$ is the subset of $\obsdata$ restricted to the descendants of node  $\node{\taxaI}$.
We compute these statistics at the tips as described in Section \ref{sec:fastLikelihood} and at internal nodes as described in Section 2.1.2 of \citetsi{hassler2020inferringsup}.

Once we have computed these statistics, we draw the factors at the root from their full conditional distribution $\cdist{\facrow{2\ntaxa - 1}}{\obsdata, \load,\errprec,\tree, \rootmean,\pss}$ as described by Equation 13 in \citetsi{hassler2020inferringsup}.
After sampling the factors $\facrow{\nroot}$ at the root node $\node{\nroot}$ from their full conditional distribution, we perform a pre-order traversal of the tree sampling from $\cdist{\facrow{\taxaI}}{\facrow{\parent{\taxaI}}, \obsdatabelow{\taxaI}, \load, \errprec, \tree}$ for $\traitI = 1\too2\ntaxa - 2$ as described in Section 2.2.1 of \citetsi{hassler2020inferringsup}.
After we have completed this pre-order traversal, we have sampled from the full conditional distribution of $\fac = \left(\facrow{1}\too\facrow{\ntaxa}\right)\trans$.

\subsection{Conjugate Gibbs Sampler on the Loadings \texorpdfstring{$\load$}{L}}\label{ap:loadGibbs}

Here we describe our procedure for sampling from $\cdist{\load}{\obsdata,\fac,\errprec}$ via the conjugate Gibbs sampler developed by \citetsi{lopes2004bayesiansup} and \citetsi{tolkoff2017phylogeneticsup}.
Let us first introduce notation related to both structured sparsity in the loadings and missing data.
Let the $\nfac$-dimensional vector $\loadcol{\traitI}$ and $\ntaxa$-dimensional vector $\datacol{\traitI}$ be the $\traitI^{\text{th}}$ column of $\load$ and $\data$ respectively for $\traitItoo$.
Let $\nzinds{\traitI} \subseteq \{1\too\nfac\}$ be the indices corresponding to the unconstrained elements of $\loadcol{\traitI}$ (i.e.~those that are not fixed at 0), and let $\nminds{\traitI}\subseteq \{1\too\ntaxa\}$ be the indices of the observed (non-missing) elements of $\datacol{\traitI}$.
Finally let the sub-vectors $\loadcolnz{\traitI}$ and $\facrownz{\taxaI}{\traitI}$ be the elements of $\loadcol{\traitI}$ and $\facrow{\taxaI}$, respectively, restricted to the indices in $\nzinds{\traitI}$, and let $\datacolnm{\traitI}$ be the elements of $\datacol{\traitI}$ restricted to the elements in $\nminds{\traitI}$ for $\taxaItoo$ and $\traitItoo$.
Note that conditional on the latent factors, the full conditional distributions of each column of the loadings are independent.
Additionally, the full conditional of $\loadcol{\traitI}$ depends only on $\datacol{\traitI}$, and does not depend on the other columns of the data matrix $\data$ \citepsi{lopes2004bayesiansup}.
As such, we draw from $\cdist{\loadcolnz{\traitI}}{\fac, \datacolnm{\traitI}, \errprec}$ as follows:

\begin{equation}\label{eq:loadGibbs}
\begin{aligned}
\cdens{\loadcolnz{\traitI}}{\datacolnm{\traitI}, \fac, \errprecel{\traitI}} &\propto \cdens{\datacolnm{\traitI}}{\loadcolnz{\traitI}, \load, \errprecel{\traitI}} \dens{\loadcolnz{\traitI}}\\
&= \prod_{\taxaI \in \nminds{\traitI}}\cdens{\datum{\taxaI\traitI}}{\facrow{\taxaI},\loadcolnz{\traitI},\errprecel{\traitI}}\dens{\loadcolnz{\traitI}} \\
&= \prod_{\taxaI \in \nminds{\traitI}} \ndens{\datum{\taxaI\traitI}}{\loadcolnz{\traitI}\trans\facrownz{\taxaI}{\traitI}}{\errprecel{\traitI}} \ndens{\loadcolnz{\traitI}}{\loadmean}{\loadprec{\traitI}} \\
&= \ndens{\loadcolnz{\traitI}}{\loadfcpmean{\traitI}}{\loadfcpprec{\traitI}}
\end{aligned}
\end{equation}
where $\loadprec{\traitI} = 1 / \loadVariid \I{\cardinality{\nzinds{\traitI}}}$, $\loadfcpprec{\traitI} = \loadprec{\traitI} + \errprecel{\traitI}\sum_{\taxaI \in \nminds{\traitI}} \facrownz{\taxaI}{\traitI}\facrownz{\taxaI}{\traitI}\trans$ and $\loadfcpmean{\traitI} = \loadfcpprec{\traitI}\inv\left(\loadprec{\traitI}\loadmean + \errprecel{\traitI}\sum_{\taxaI \in \nminds{\traitI}} \datum{\taxaI\traitI} \facrownz{\taxaI}{\traitI}\right)$.

Computing $\loadfcpprec{\traitI}$ has computational complexity $\bigo{\ntaxa\nfac^2}$, so computing all $\ntraits$ precisions has overall complexity $\bigo{\ntaxa\ntraits\nfac^2}$.
Once the precisions have been computed, computing the means has complexity $\bigo{\ntaxa\ntraits\nfac + \ntraits\nfac^3}$, which contributes relatively little to overall computation time as $\ntaxa >> \nfac$ for most problems.
Note that if the data are completely observed and there is no structured sparsity in the loadings, then $\loadfcpprec{\traitI} = \loadprec{\traitI} + \errprecel{\traitI}\fac\trans\fac$.
In that case, we only need to compute $\fac\trans\fac$ once (not $\ntraits$ times), which brings the overall complexity down to $\bigo{\ntaxa\ntraits\nfac}$ (as we still need to compute the means for al $\ntraits$ columns of $\load$).
Drawing all $\loadcol{\traitI}$ for $\traitItoo$ results in a complete sample from the full conditional distribution of $\load$.

\section{Loadings Gradient Calculation}\label{ap:gradCalc}
The fact that $\cdist{\data}{\fac, \load} \sim \mndist{\load\fac}{\I{\ntaxa}}{\errprec\inv}$ implies
\begin{equation}
\begin{aligned}
\grad{\cdens{\data}{\fac, \load}}{\load} &= \gradvar{\load}\left[ \left(2\pi\right)^{-\ntaxa\ntraits/2} \detOp{\errprec}^{\ntaxa/2}\detOp{\I{\ntaxa}}^{-\ntraits/2}\right.\\
&\hspace{2cm}\times\left.\expOp{-\frac{1}{2} \trOp{\errprec\left(\fac\load - \data\right)\trans \I{\ntaxa}\inv \left(\fac\load - \data\right)}} \right] \\
&= \left(2\pi\right)^{-\ntaxa\ntraits/2} \detOp{\errprec}^{\ntaxa/2} \gradvar{\load}\left[\expOp{-\frac{1}{2} \trOp{\errprec\left(\fac\load - \data\right)\trans\left(\fac\load - \data\right)}} \right]\\
&=  \left(2\pi\right)^{-\ntaxa\ntraits/2} \detOp{\errprec}^{\ntaxa/2} \left(\expOp{-\frac{1}{2} \trOp{\errprec\left(\fac\load - \data\right)\trans\left(\fac\load - \data\right)}} \right)\\
&\hspace{2cm} \times \gradvar{\load} \left[ -\frac{1}{2} \trOp{\errprec\left(\fac\load - \data\right)\trans\left(\fac\load - \data\right)}\right] \\
&= -\frac{1}{2} \cdens{\data}{\fac, \load} \gradvar{\load} \trOp{\errprec\left(\fac\load - \data\right)\trans\left(\fac\load - \data\right)}\\
&= -\frac{1}{2} \cdens{\data}{\fac, \load} \left[ \gradvar{\load}\trOp{\errprec\load\trans\fac\trans\fac\load} - 2\gradvar{\load} \trOp{\errprec\data\trans\fac\load} + \gradvar{\load}\trOp{\errprec\data\trans\data}\right] \\
&= -\frac{1}{2} \cdens{\data}{\fac, \load}\left[2\fac\trans\fac\load\errprec - 2\fac\trans\data\errprec + \bzero\right] \\
&= \cdens{\data}{\fac, \load} \left[\fac\trans\data\errprec - \fac\trans\fac\load\errprec\right]
\end{aligned}
\end{equation}

\section{Post-Processing Procedure} \label{ap:svd}

We employ singular value decomposition (SVD) to enforce the orthogonality constraint on the loadings via post-processing.
\def\facSample{\fac\upstate}
\def\facOrthoSample{{\fac^{\perp}}\upstate}
In practice, we sample from the orthogonally-constrained loadings as follows.
Let $\loadsample$ be a sample from the posterior distribution $\cdist{\load}{\data}$ at the $\state^{\text{th}}$  state in the MCMC chain.
For each $\loadsample$, we compute the SVD $\loadsample = \svdUsample\svdSsample{\svdVsample}$ where $\svdUsample$ is a $\nfac\times\nfac$ orthonormal matrix and $\svdSsample$ and $\svdVsample$ retain their constraints from Section \siref{sec:shrink} (i.e.~$\svdSsample$ is diagonal with descending positive entries and $\svdVsample{\svdVsample}\trans = \I{\nfac}$).
While the parameter $\svdU$ is not identifiable, $\svdS$ and $\svdV$ are \citepsi{holbrook2016bayesiansup}.
As such, we then treat $\loadOrthoFullSample = \svdSsample{\svdVsample}$ as (now identifiable) samples from the posterior of the loadings.
If we also sample the factors $\fac$, we rotate the factors to sample from $\facOrthoSample = \facSample\svdUsample$ to ensure that $\facOrthoSample\loadOrthoFullSample = \facSample\svdUsample \svdSsample{\svdVsample} = \facSample\loadsample$.

\section{Loadings Scale Full Conditional Distribution}\label{sec:scaleGibbsDetails}

We detail our derivation of Equation \siref{eq:scaleGibbs} below.
Recall that we define the $\nfac$-vector $\loadScales$ such that $\loadScale = \diagOp{\loadScales}$, and note that all proportional symbols imply log-proportional:
\begin{equation}\label{eq:scaleGibbsDetails}
\begin{aligned}
\logcdens{\loadScales}{\obsdata,\fac,\loadOrtho,\errprec}\hspace{-2cm}&\\
&\propto \logcdens{\obsdata}{\loadScales, \fac,\loadOrtho,\errprec} + \scalePrior \\
&= \sumTraits \logcdens{\obsdatacol{\traitI}}{\loadScales, \fac, \orthocoli,\errpreci} + \scalePrior \\
&\propto -\frac{1}{2}\sumTraits\errpreci\errTrait\trans\taxamismati\errTrait + \scalePrior \\
&\propto -\frac{1}{2}\sumTraits\errpreci\left(\orthocoli\trans\loadScale\fac\trans\taxamismati\fac\loadScale\orthocoli - 2\orthocoli\trans\loadScale\fac\trans\taxamismati\obsdatacol{\traitI}\right) + \scalePrior \\
&\propto -\frac{1}{2}\sumTraits\errpreci\left(\loadScales\trans\diagOp{\orthocoli}\fac\trans\taxamismati\fac\diagOp{\orthocoli}\loadScales - 2\loadScales\trans\diagOp{\orthocoli}\fac\trans\taxamismati\obsdatacol{\traitI}\right) + \scalePrior \\
&\propto-\frac{1}{2}\loadScales\trans\left(\sumTraits\errpreci\diagOp{\orthocoli}\fac\trans\taxamismati\fac\diagOp{\orthocoli}\right)\loadScales \\
&\hspace{0.5in} -\loadScales\trans\left(\sumTraits\errpreci\diagOp{\orthocoli}\fac\trans\taxamismati\obsdatacol{\traitI}\right) + \scalePrior\\
&\propto -\frac{1}{2}\loadScales\trans\left(\scalePostPrecLong\right)\loadScales\\
&\hspace{0.5in} -\loadScales\trans\left(\scalePostMeanSub\right) \\
&\propto  -\frac{1}{2}\left(\loadScales - \scalePostMean\right)\trans\scalePostPrec\left(\loadScales - \scalePostMean\right),
\end{aligned}
\end{equation}
where
\begin{equation}
\begin{aligned}
\scalePostPrec &= \scalePostPrecLong\text{ and} \\
\scalePostMean &= \scalePostPrec\inv\left(\scalePostMeanSub\right)
\end{aligned}
\end{equation}
This implies
\begin{equation}
\logcdens{\loadScales}{\obsdata,\fac,\loadOrtho,\errprec} = \ndens{\loadScales}{\scalePostMean}{\scalePostPrec}.
\end{equation}

\section{Sign Constraint on the Loadings} \label{ap:signConstraint}

\def\maxInd{\traitI_\mathrm{max}}
\def\sumStates{\sum_{\state = 1}^\nstates}
\def\absStateMean{\bar{\loadel{}}_{\traitI\facI}^\mathrm{abs}}
Regardless of which prior (i.i.d.~vs~shrinkage) or constraint (sparsity vs orthogonality) we choose, we must enforce a sign constraint on a single element in each row of $\load$ for full identifiability.
Let $\signindsi\in\{1\too\ntraits\}$ be the index of the $\nfac^{\text{th}}$ row of $\load$ with the sign constraint (i.e.~require $\loadel{\signindsi\facI} \geq 0$).
If the sample $\loadel{\facI\signindsi}\upstate < 0$, then we simply multiply row $\facI$ of $\loadsample$ by $-1$ to ensure $\loadel{\facI\signindsi}\upstate \geq 0$.
These $\nfac$ sign-constrained elements are not required to be in the same row of $\load$, and we choose these rows in a way that maximizes the posterior identifiability of $\load$.
In practice, we apply a simple heuristic where for $\facItoo$
\begin{equation}
\signindsi = \argmaxOp{\traitI\in 1\too\ntraits}{ \frac{\absStateMean}{\sqrt{\sumStates \parens*{\absOp{\loadelsample{\traitI\facI}} - \absStateMean}^2}}}\quad\text{and}\quad \absStateMean = \frac{1}{\nstates}\sumStates \absOp{\loadelsample{\traitI\facI}}.
\end{equation}
In the absence of sign constraints, the marginal posteriors of many elements of $\load$ are bimodal and symmetric across zero.
Our heuristic aims to find an index in each column of $\load$ with low mass near 0 and simply chose the positive mode.

\section{Sampling from \texorpdfstring{$\errprec$}{Lambda}} \label{ap:precInference}

Regardless of the prior on the loadings, we sample from $\cdist{\errprec}{\fac, \obsdata, \load}$ using the same conjugate Gibbs sampler as \citetsi{tolkoff2017phylogeneticsup} in conjunction with the data augmentation algorithm from Section \siref{sec:gibbsLoadings}.
The $\gammaDist{\errshape}{\errrate}$ (shape, rate parameterization) prior on the diagonal elements of $\errprec$ results in a simple expression for the full conditional distribution of $\errprecel{\traitI}$ for $\traitI = 1\too\ntraits$ conditional on the factors $\fac$.
Specifically, each $\errprecel{\traitI}$ is distributed as
\begin{equation}
\cdist{\errprecel{\traitI}}{\obsdata, \fac, \load} \sim \gammadist{\errshape + \frac{\ntaxaobs{\traitI}}{2}}{\errrate + \frac{1}{2}\sum_{\taxaI = 1}^\ntaxa \misind{\taxaI\traitI}\left(\datum{\taxaI\traitI} - \loadcol{\traitI}\trans\facrow{\taxaI}\right)^2}.
\end{equation}
This computation only requires run time $\bigo{\ntaxa\ntraits\nfac}$ and, in our experience, time spent estimating $\errprec$ does not contribute significantly to the overall run time of the MCMC chain.

Note that as with the loadings in Section \siref{sec:hmcLoadings}, we also derive a strategy for sampling from these precisions without conditioning on $\fac$ via HMC.
As we are satisfied with the \citetsi{tolkoff2017phylogeneticsup} procedure, we have not implemented this strategy, but the derivation can be found below.
Naturally, this HMC sampler requires we compute the gradient of the likelihood with respect to the loadings as follows:
\def\precExp{\frac{\ntaxaobs{\traitI}}{2}\errprecel{\traitI}\inv - \frac{1}{2} \left(\fac\loadcol{\traitI} - \obsdatacol{\traitI}\right)\trans \taxamismat{\traitI}\left(\fac\loadcol{\traitI} - \obsdatacol{\traitI}\right)}
\begin{equation}
\begin{aligned}
\pderiv{\log\cdens{\obsdata}{\errprec}}{\errprecel{\traitI}} &= \frac{1}{\cdens{\obsdata}{\errprec}} \int \dens{\fac} \pderiv{\cdens{\obsdata}{\fac, \errprec}}{\errprecel{\traitI}} \diff\fac \\
&= \frac{1}{\cdens{\obsdata}{\errprec}} \int \dens{\fac} \cdens{\obsdata}{\fac, \errprec} \\ &\hspace{2cm}\times \left( \precExp \right)\diff\fac \\
&= \cexpectOp{\precExp}{\obsdata, \errprec} \\
&= \frac{\ntaxaobs{\traitI}}{2}\errprecel{\traitI}\inv -\frac{1}{2}\loadcol{\traitI}\trans\cexpectOp{\fac\trans\taxamismat{\traitI}\fac}{\obsdata, \errprec}\loadcol{\traitI} +\loadcol{\traitI}\trans\cexpectOp{\fac\trans}{\obsdata, \errprec}\taxamismat{\traitI}\obsdatacol{\traitI}\\
&\quad\quad-\frac{1}{2} {\obsdatacol{\traitI}}\trans\taxamismat{\traitI}\obsdatacol{\traitI}
\end{aligned}
\end{equation}
The conditional expectations of the latent factors are the same as in Section \siref{sec:hmcLoadings}.
Note that we restrict $\errprec$ to be diagonal, so we only consider the diagonal elements of the gradient.
Once we have computed this gradient, we employ it in standard HMC to sample from the full conditional of $\errprec$.

\section{Timing}

\subsection{Simulation Details}\label{ap:timing}

To simulate each data set for the timing comparison, we generate a random coalescent tree with $\ntaxa$ tips \citepsi{kingman1982coalescentsup}.
We then simulate the factors $\fac$ according to $\nfac$ independent Brownian diffusion processes on the tree and subsequently re-scale the factors so that each column has unit variance.
We draw $\loadOrtho$ from a uniform distribution on the Stiefel manifold.
To avoid identifiability challenges associated with values of $\loadScale$ having similar magnitudes, we set $\loadScaleEl{\facI} = 2^{-\facI}\sqrt{\ntraits}$ for $\facItoo$.
Note that we multiply by $\sqrt{\ntraits}$ so that the expectations of $\loadel{\facI\traitI}^2 = \loadScaleEl{\facI}^2\loadOrthoEl{\facI\traitI}^2$ remain the same regardless of $\ntraits$.
We sample the residual variances $\errprecel{\traitI}\inv$ independently from $\gammaDist{2}{4}$ for $\traitItoo$, which keeps the contribution of the residual variance to the total variance similar to that of the latent factors.
Finally, we draw $\error\sim\mnDist{\mathbf{0}}{\I{\ntaxa}}{\errprec\inv}$ and compute $\data = \fac\loadScale\loadOrtho + \error$.
As all methods rely on the same principles for handling missing data, we do not remove any observations from the simulated data sets.

When performing inference, we assume the tree is fixed to its true value used to simulate the factors $\fac$.
We use the orthogonality constraint on the loadings and employ the post-processing regime discussed in Section \siref{sec:svdProcessing} to rotate results from each sampler (except the one associated with the orthogonal shrinkage prior) to enforce this constraint.
For the model with the orthogonal shrinkage prior, we assume both forced ordering and spacing ($\scaleThreshold = 0.9$).

\subsection{Effective Sample Size Calculations} \label{ap:ess}

\def\essSymb{\text{ESS}}
\def\simi{m}
\def\ess{{\essSymb}_{\facI\traitI}^{(\simi)}}
\def\essmin{\essSymb^{(\simi)}_\mathrm{min}}
\def\essmeanmin{\overline{\essSymb}_{\mathrm{min}}}
\def\simtime{t^{(\simi)}}

To understand the relative performance of each inference regime, we compare the effective sample size (ESS) per second of the loadings across all four samplers.
Draws from an MCMC simulation are often auto-correlated, and the total number of steps in the chain is rarely a direct proxy for our confidence in the posterior estimates.
ESS approximates the number of \emph{independent} samples from the chain.
As researchers typically set a minimum ESS threshold to determine the length of MCMC simulations, we compare the minimum ESS per unit time.
Let $\ess$ be the effective sample size for $\loadij$ in replicate $\simi$ and $\essmin = \min_{\facI,\traitI}\ess$ for $\simi=1\too\nsims$. %
We compute $\essmeanmin = \frac{1}{\nsims}\sum_{\simi=1}^{\nsims}\essmin / \simtime$ for all models, where $\simtime$ is the time required for the $\simi^{\text{th}}$ MCMC simulation.
Actual ESS values were calculated using the Julia package MCMCDiagnostics.jl.
We compare these values in Figure \siref{fig:timing} and \siTab~\ref{tb:timing}.%

\begin{table}[H]
	\centering
	\caption{Comparison of computational efficiency. Effective sample size computed using the Julia package MCMCDiagnostics.jl.}
	\label{tb:timing}
	\small
	\begin{tabular}{rrrrrrrrrr}\toprule
		\multirow{2}{*}{$\ntaxa$} & \multirow{2}{*}{$\ntraits$} & \multirow{2}{*}{$\nfac$} & \multicolumn{4}{c}{minimum ESS per minute} & \multicolumn{3}{c}{speed increase over sampled} \\\cmidrule(l{2pt}r{2pt}){4-7}\cmidrule(l{2pt}r{2pt}){8-10}
		&& & Sampled & Gibbs & HMC & orthogonal & Gibbs & HMC & orthogonal\\ \midrule
		\multirow{9}{*}{50} & \multirow{3}{*}{10} & 1 & 530 & 5100 & 5700 & 2000 & 9.8$\times$ & 11.0$\times$ & 3.8$\times$ \\
		&  & 2 & 500 & 3900 & 2500 & 810 & 7.8$\times$ & 4.9$\times$ & 1.6$\times$ \\
		&  & 4 & 680 & 2200 & 1400 & 450 & 3.3$\times$ & 2.0$\times$ & 0.7$\times$ \\\cmidrule{2-10}
		& \multirow{3}{*}{100} & 1 & 190 & 1400 & 1700 & 170 & 7.6$\times$ & 9.1$\times$ & 0.89$\times$ \\
		&  & 2 & 150 & 1000 & 870 & 130 & 7.1$\times$ & 5.9$\times$ & 0.89$\times$ \\
		&  & 4 & 52 & 550 & 250 & 20 & 11$\times$ & 4.7$\times$ & 0.39$\times$ \\\cmidrule{2-10}
		& \multirow{3}{*}{1000} & 1 & 34 & 460 & 250 & 5.2 & 14$\times$ & 7.4$\times$ & 0.15$\times$ \\
		&  & 2 & 27 & 390 & 85 & 0.87 & 14$\times$ & 3.1$\times$ & 0.032$\times$ \\
		&  & 4 & 23 & 320 & 23 & 0.51 & 14$\times$ & 1.0$\times$ & 0.022$\times$ \\\cmidrule{1-10}
		\multirow{9}{*}{100} & \multirow{3}{*}{10} & 1 & 270 & 4100 & 3000 & 1100 & 15$\times$ & 11$\times$ & 4.0$\times$ \\
		&  & 2 & 160 & 2100 & 2000 & 400 & 13$\times$ & 12$\times$ & 2.5$\times$ \\
		&  & 4 & 51 & 680 & 500 & 110 & 13$\times$ & 9.9$\times$ & 2.1$\times$ \\\cmidrule{2-10}
		& \multirow{3}{*}{100} & 1 & 33 & 360 & 480 & 94 & 11$\times$ & 14$\times$ & 2.9$\times$ \\
		&  & 2 & 18 & 240 & 290 & 35 & 13$\times$ & 16$\times$ & 1.9$\times$ \\
		&  & 4 & 17 & 200 & 83 & 38 & 12$\times$ & 4.8$\times$ & 2.2$\times$ \\\cmidrule{2-10}
		& \multirow{3}{*}{1000} & 1 & 3.9 & 54 & 53 & 2.9 & 14$\times$ & 14$\times$ & 0.75$\times$ \\
		&  & 2 & 2.5 & 82 & 15 & 0.98 & 33$\times$ & 5.8$\times$ & 0.39$\times$ \\
		&  & 4 & 2.0 & 99 & 5.3 & 0.19 & 49$\times$ & 2.6$\times$ & 0.092$\times$ \\\cmidrule{1-10}
		\multirow{9}{*}{500} & \multirow{3}{*}{10} & 1 & 5.0 & 740 & 460 & 170 & 150$\times$ & 92$\times$ & 33$\times$ \\
		&  & 2 & 3.4 & 260 & 280 & 59 & 77$\times$ & 83$\times$ & 17$\times$ \\
		&  & 4 & 1.7 & 160 & 170 & 30 & 93$\times$ & 98$\times$ & 18$\times$ \\\cmidrule{2-10}
		& \multirow{3}{*}{100} & 1 & 0.77 & 95 & 110 & 25 & 120$\times$ & 140$\times$ & 32$\times$ \\
		&  & 2 & 0.37 & 20 & 28 & 5.4 & 56$\times$ & 77$\times$ & 15$\times$ \\
		&  & 4 & 0.46 & 18 & 12 & 3.7 & 40$\times$ & 25$\times$ & 8.1$\times$ \\\cmidrule{2-10}
		& \multirow{3}{*}{1000} & 1 & 0.02 & 1.8 & 0.71 & 0.68 & 90$\times$ & 35$\times$ & 34$\times$ \\
		&  & 2 & 0.018 & 2.4 & 0.65 & 0.11 & 130$\times$ & 36$\times$ & 6.1$\times$ \\
		&  & 4 & 0.011 & 1.5 & 0.16 & 0.032 & 140$\times$ & 15$\times$ & 2.9$\times$ \\\cmidrule{1-10}
		\multirow{9}{*}{1000} & \multirow{3}{*}{10} & 1 & 1.1 & 170 & 290 & 58 & 160$\times$ & 270$\times$ & 54$\times$ \\
		&  & 2 & 0.54 & 84 & 190 & 28 & 160$\times$ & 350$\times$ & 52$\times$ \\
		&  & 4 & 0.24 & 49 & 80 & 10 & 210$\times$ & 340$\times$ & 44$\times$ \\\cmidrule{2-10}
		& \multirow{3}{*}{100} & 1 & 0.098 & 35 & 38 & 9.2 & 350$\times$ & 390$\times$ & 94$\times$ \\
		&  & 2 & 0.064 & 15 & 12 & 2.8 & 230$\times$ & 180$\times$ & 44$\times$ \\
		&  & 4 & 0.065 & 7.6 & 5.8 & 1.0 & 120$\times$ & 90$\times$ & 15$\times$ \\\cmidrule{2-10}
		& \multirow{3}{*}{1000} & 1 & 0.0017 & 0.5 & 0.25 & 0.3 & 300$\times$ & 150$\times$ & 180$\times$ \\
		&  & 2 & 0.0015 & 0.67 & 0.15 & 0.085 & 450$\times$ & 100$\times$ & 57$\times$ \\
		&  & 4 & 0.0015 & 0.4 & 0.06 & 0.02 & 270$\times$ & 40$\times$ & 14$\times$ \\ \bottomrule
	\end{tabular}
\end{table}

\section{Phylogenetic Latent Liability Model} \label{ap:latentLiability}
\newcommand*{\proposal}[1]{{#1}}
\def\latlproposed{\proposal{\latlel{}{}}_{\taxaI\traitI}}

In the case of binary traits, we assume the latent liability model of \citetsi{cybis2015assessingsup}.
Specifically, rather than assuming the observations $\data = \fac\load + \error$, we introduce an additional latent variable $\latl = \left\{\latlij\right\}$ for $\taxaItoo$, $\traitItoo$ and assume $\latl = \fac\load + \error$.
These latent liabilities $\latlij$ are connected to the observations $\datumij$ via the link function $\datumij = \latlink{\traitI}{\latlij}$ where $\latlink{\traitI}{\arbarg} = \arbarg$ if trait $\traitI$ is continuous, $\latlink{\traitI}{\arbarg} = \Indicator{\arbarg \leq 0}$ if $\traitI$ is binary.

Under this model, the full conditional distributions of the latent liabilities are independent truncated Gaussian distributions with densities
\begin{equation}
\cdens{\latlij}{\datumij, \facrow{\taxaI}, \loadcol{\traitI}, \errprecel{\traitI}, \thresholds}\sim\ndens{\latlij}{\facrow{\taxaI}\trans\loadcol{\traitI}}{\errprecel{\traitI}}\Indicator{\latlink{\traitI}{\latlij} = \datumij}.
\end{equation}
As these full conditional distributions are independent, we can sample from them efficiently via a simple rejection sampler. %
Specifically, we first draw from $\cdist{\fac}{\latl, \errprec, \tree}$ as in Section \siref{sec:gibbsLoadings}.
We then sample the proposal $\latlproposed \sim \ndist{\facrow{\taxaI}\trans\loadcol{\traitI}}{1 / \errprecel{\traitI}}$ that we accept if $\latlink{\traitI}{\latlij} = \datumij$ and reject otherwise.
Note that for each discrete trait $\traitI$, we must also fix $\errprecel{\traitI} = 1$ to ensure the variance of the latent traits $\traitI$ are identifiable \citepsi[see][]{tolkoff2017phylogeneticsup}.

\section{Phylogenetic Tree Inference} \label{ap:treeInference}

\subsection{Yeast Phylogeny} \label{ap:yeastPhylo}
For they yeast analysis, we first infer a phylogenetic tree for the 154 phenotyped strains using the 2.8 megabase DNA sequence alignment of \citetsi{gallone2016domesticationsup} (see subsection \emph{Phylogenetic Tree for the Sequenced Collection} in \emph{Methods} of \citetsi{gallone2016domesticationsup} for details).
Our phylogenetic tree model includes an uncorrelated relaxed clock model \citepsi{drummond2006relaxedsup}, an HKY+G substitution model \citepsi{hasegawa85sup,yang1994maximumsup} and a constant-population coalescent prior on the tree \citepsi{kingman1982coalescentsup}.

We perform MCMC simulation via BEAST \citepsi{BEASTsup} to approximate the posterior distribution of the phylogenetic tree.
We run the MCMC chain for 10 million states, sampling the tree and related parameters every thousand states and the factor related parameters every 10 thousand states.
Inspection of relevant trace plots indicated the the MCMC chain had achieved stationarity by 1 million states, and we exclude the first million states as burn-in.
We compute the maximum clade credibility (MCC) tree as a point estimate of the phylogenetic tree using TreeAnnotator \citepsi{treeAnnotatorsup}. %

\subsection{New World Monkey Phylogeny} \label{ap:nwmTree}

We simultaneously infer the NWM tree structure with the latent factor model using DNA sequence alignments of \citetsi{aristide2015modelingsup}.
To infer the tree structure, we partition the taxa into four monophyletic clades consisting of the 1) \emph{Atelidae}, 2) \emph{Aotidae} and \emph{Callitrichidae}, 3) \emph{Cebidae} and 4) \emph{Pitheciidae} respectively and place zero prior probability on tree topologies that do not maintain these clades.
Otherwise, we use the same phylogenetic tree model and inference procedure as described in \siSec~\ref{ap:yeastPhylo}.

\bibliographystylesi{chicago}
\bibliographysi{pfa_sup}

\end{document}